\documentclass[aps,prl,twocolumn,footinbib,notitlepage]{revtex4-1}
\usepackage{graphicx}
\usepackage{amsmath,amsthm,amssymb,mathtools}
\usepackage{color}
\usepackage[none]{hyphenat}

\def\blue#1{{\color{blue}#1}}
\usepackage{tcolorbox}
\usepackage{cancel}
\usepackage{braket}
\usepackage[pdftex,colorlinks=true,linkcolor=Cerulean,citecolor=Cerulean,urlcolor=Cerulean]{hyperref}
\hypersetup{linkcolor=blue,citecolor=blue,urlcolor=blue}
\hypersetup
{
  pdfauthor={J. Enrique V\'azquez\,-Lozano (juavazlo@ntc.upv.es) \& Alejandro Mart\'inez (amartinez@ntc.upv.es)},
  pdfsubject={Nanophotonics Technology Center -- Universitat Polit\`ecnica de Val\`encia},
  pdftitle={Optical Chirality in Dispersive and Lossy Media (main text+supp mat)},
  pdfkeywords={optical chirality, conservation laws, continuity equation, dispersive and lossy (lossless) medium, Lorentz medium, nanophotonics}
}

\begin{document}
\title{Optical Chirality in Dispersive and Lossy Media}
\author{J. Enrique V\'azquez\,-Lozano}
\email{juavazlo@ntc.upv.es}
\affiliation{Nanophotonics Technology Center, Universitat Polit\`ecnica de Val\`encia, Camino de Vera s/n, 46022 Valencia, Spain}
\author{Alejandro Mart\'inez}
\email{amartinez@ntc.upv.es}
\affiliation{Nanophotonics Technology Center, Universitat Polit\`ecnica de Val\`encia, Camino de Vera s/n, 46022 Valencia, Spain}
\date{\today}

\begin{abstract}
Several dynamical properties such as energy, momentum, angular momentum, and optical helicity have been recently reexamined in dispersive and lossless media. Here, we address a parallel derivation for the optical chirality, extending it so as to include dissipative effects as well. To this end, we first elaborate on the most complete form of the conservation law for the optical chirality, without any restrictions on the nature of the medium. As a result we find a general expression for the optical chirality density both in lossless and lossy dispersive media. Our definition is perfectly consistent with that originally introduced for electromagnetic fields in free space, and is applicable to any material system, including dielectrics, plasmonic nanostructures, and left-handed metamaterials.
\end{abstract}
\maketitle
\sloppy

\textit{Introduction.\textemdash}Local dynamical properties such as energy, linear momentum, and angular momentum, among others, are conserved quantities for electric and magnetic fields in vacuum \cite{Jackson,Novotny}. In fact, leaving aside the physical meaning, there exists an infinity class of conserved quantities for free electromagnetic fields \cite{Kibble1965,Fushchich1992}. In particular, in 1964 Lipkin demonstrated the existence of a set containing ten new independent conservation laws for electromagnetic radiation in vacuum \cite{Lipkin1964}. Originally, these tensorial quantities were merely conceived as mathematical entities theretofore unknown, and having no ready physical significance. That is why they were collectively referred to as the $ij$\textit{-zilches} (which literally means ``nothingness''), where $i$ and $j$ stand for the labels indicating the tensor indices. Since then, there have been many efforts in searching for a physically meaningful picture for these quantities \cite{Morgan1964,Candlin1965,Calkin1965}. 

Recent advances in near-field optics attempting to achieve full spatiotemporal control of light-matter interactions \cite{Brixner2005Aeschlimann2007} has led to a renewed interest in Lipkin's zilches as a measure of the handedness, or knottedness, of the streamlines describing highly contorted optical fields \cite{Yang2009}. In this regard, and motivated by the possibility for enhancing the chiroptical effects (such as circular dichroism (CD) \cite{Barron}), which leads to enantioselective signals far larger than that due to circularly polarized light (CPL), Tang and Cohen introduced the $00$-zilch as a measure of the local density of optical chirality \cite{Tang2010}:
\begin{equation}
\mathcal{C}_{\rm vacuum}\equiv\frac{1}{2}\left[\varepsilon_0\boldsymbol{\mathcal{E}}\cdot\left(\nabla\times\boldsymbol{\mathcal{E}}\right)+\mu_0\boldsymbol{\mathcal{H}}\cdot\left(\nabla\times \boldsymbol{\mathcal{H}}\right)\right],
\label{Eq1}
\end{equation}
where $\varepsilon_0$ and $\mu_0$ are the permittivity and permeability of vacuum, respectively, and $\boldsymbol{\mathcal{E}}({\bf r},t)$ and $\boldsymbol{\mathcal{H}}({\bf r},t)$ are the local, time-dependent electric and magnetic fields. Shortly after, this definition for the optical chirality was successfully used in enhanced CD spectroscopic measurements for the experimental detection and characterization of chiral biomolecules \cite{Hendry2010}, thus confirming its physical significance, and highlighting the feasibility for practical applications. The extremely high sensitivity in the chiroptical responses (enhancement factors up to $6$ orders of magnitude were reported) was attributed to the presence of the incipiently postulated \textit{superchiral fields} \cite{Tang2011}. However, as pointed out in Refs. \cite{Choi2012,Bliokh2011}, on account of the energy conservation, there should be an upper bound lowering those enhancements. In this respect, it was argued that this fundamental restriction ought to limit the enhancement factor up to two orders of magnitude \cite{Choi2012}; the other four orders should come from the highly twisted evanescent near-field modes \cite{Bliokh2011,Barr2018}. It then follows that, essentially, the main requirement for the occurrence of strengthened chiroptical influence in light-matter interaction relies on the complexity in the structure of the electromagnetic field distribution \cite{Yang2009,Kramer2017,Barr2018}. For this reason, metallic nanostructures represent ideal candidates for investigating chirality-based applications and functionalities in nanophotonics \cite{Schaferling2012,Meinzer2013,Valev2013,Nesterov2016,Collins2017,Luo2017,Hentschel2017,Schaferling}. It is certainly surprising, however, that, most of the previous studies on this issue build on the earliest definition for the optical chirality density \cite{Lipkin1964}, which is only valid for monochromatic optical fields in free space \cite{Tang2010,Tang2011,Bliokh2011}. Still, there are few works attempting to extend the definition of the optical chirality density to linear \cite{Choi2012}, gyrotropic \cite{Proskurin2017}, or dispersive lossless media \cite{Philbin2013}.

Inspired by the latest theoretical results concerning the dispersive features of the electromagnetic energy-momentum, the optical orbital and spin angular momentum \cite{Philbin2011,Philbin2012, Bliokh2017}, and the electromagnetic helicity \cite{Alpeggiani2018}, in this Letter we report on the optical chirality in lossless and lossy dispersive media. Special emphasis is placed on the role of the mathematical structure of the corresponding conservation law. Indeed, building on previous approaches addressing the electromagnetic energy density considering dispersion as well as dissipation \cite{SM}, we put forward a complete description for the optical chirality conservation law valid for arbitrarily structured optical fields. The only restriction we need to impose relies on the electromagnetic characterization of the medium, which must be fitted by Lorentzian line shapes. Hence, our results are completely general \cite{Sehmi2017}, and are applicable to any material system, including dielectrics, semiconductors, metals, as well as metamaterials, even with negative refractive index. Further, our findings are perfectly consistent with the ones so far established for optical fields in free space \cite{Tang2010,Bliokh2011}, thus showing its physical meaningfulness.

\textit{Conservation law for the optical chirality.\textemdash} Conservation laws and symmetry properties of a physical system are, arguably, among the most important cornerstones of modern physics \cite{Fushchich}. More in-depth insights would require the standard Lagrangian formulation \cite{Soper}. Indeed, appealing to the principle of least action, the Noether's theorem states that, in the absence of sources (or sinks), conserved quantities and symmetries can be regarded as equivalent features \cite{Noether1918}. These theoretical concepts are mathematically described via continuous or discrete symmetry groups, which are in turn related to the corresponding physical transformations \cite{Tung}. Typical examples of continuous symmetries lead to the conservation of energy, linear momentum and angular momentum, which are associated with the invariance under the universal space-time transformations, i.e., translations and rotations. An insightful picture of the conserved quantities, reminiscent of the quantum formalism \cite{Weinberg}, allows one to deal with the conserved quantities as differential operators representing the generators of the corresponding infinitesimal symmetry transformations. For the above dynamical properties, the generators simply involve first derivatives with respect to the space-time coordinates acting on the electromagnetic fields, and are given explicitly by $\left\{i\partial_t,i\nabla\right\}$, for the space-time translations \cite{Philbin2011}, and $i\left({\bf r}\times\nabla\right)$, for the spatial rotations \cite{Philbin2012}. Furthermore, it was recently demonstrated that the conservation of the optical chirality is underpinned by $i\left(\partial_t\nabla\times\right)$ \cite{Philbin2013}, which must be applied on the vector potentials rather than on the electromagnetic fields. Importantly, these generators can be used to find the eigenstates of the aforementioned conserved quantities. In this regard, just as the plane waves are the eigenstates of the energy-momentum differential operator, it has been shown that the corresponding eigenstates associated to the optical chirality are the circularly polarized plane waves \cite{Philbin2013}.

\begin{figure*}[t!]
	\includegraphics[width=1\linewidth]{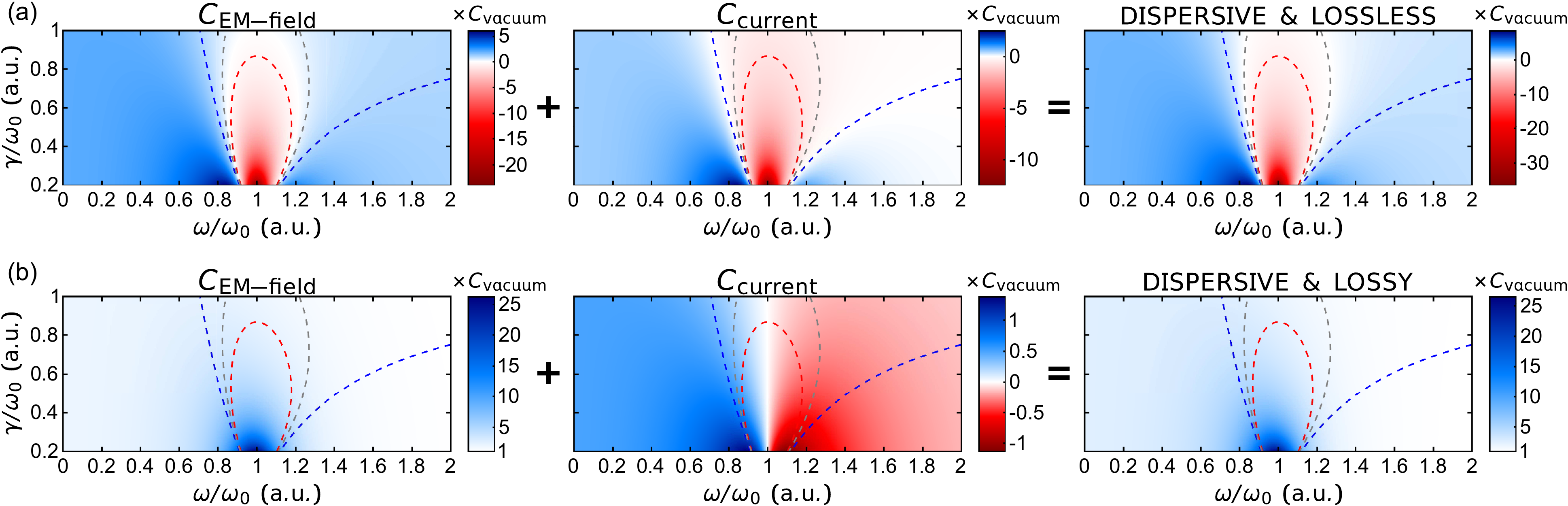} 
	\caption{Optical chirality density in (a) lossless and (b) lossy dispersive media. Material parameters correspond to a nonmagnetic medium ($\mu=1$) whose permittivity is described by a single Lorentz pole with $\omega_{\rm p}=\omega_0$. In all the cases, red, gray and blue dashed lines indicate the curves where the total contribution of the optical chirality in the lossless case is $-1$, $0$ and $1$, respectively.}
	\label{fig1}
\end{figure*}
The above scheme for identifying continuous conserved quantities only holds in the absence of sources. In the presence of charges and/or currents, conservation laws are to be expressed through the continuity equations \cite{Nienhuis2016,Corbaton2017}. Within the electromagnetic field theory, the most well-known example is perhaps the Poynting's theorem, accounting for the energy conservation \cite{Jackson,Novotny}:
\begin{equation}
\nabla\cdot \boldsymbol{\mathcal{S}}=\!-\!\left[\boldsymbol{\mathcal{E}}\cdot \partial_t \boldsymbol{\mathcal{D}}+\boldsymbol{\mathcal{H}}\cdot \partial_t\boldsymbol{\mathcal{B}}+\boldsymbol{\mathcal{J}}\cdot \boldsymbol{\mathcal{E}}\right],
\label{Eq2}
\end{equation}
where $\boldsymbol{\mathcal{S}}\equiv \boldsymbol{\mathcal{E}}\times \boldsymbol{\mathcal{H}}$ is the Poynting vector, which represent the energy flux density, and $\boldsymbol{\mathcal{D}}$, $\boldsymbol{\mathcal{B}}$, and $\boldsymbol{\mathcal{J}}$ are the time-dependent electric displacement, magnetic induction, and electric current density (or charge flux), respectively. This expression is generally valid, and can be readily obtained by taking the divergence of the energy flux density. Likewise, we can derive the time-dependent conservation law for the optical chirality from the corresponding chirality flux density \cite{Lipkin1964,Tang2010,Bliokh2011}:
\begin{equation}
\boldsymbol{\mathcal{F}}\equiv\left[\boldsymbol{\mathcal{E}}\times\left(\nabla\times \boldsymbol{\mathcal{H}}\right)-\boldsymbol{\mathcal{H}}\times\left(\nabla\times \boldsymbol{\mathcal{E}}\right)\right]/2.
\label{Eq3}
\end{equation}
With the aid of the structural Maxwell's equations and the vector identity, $\nabla\cdot\left(\boldsymbol{\mathcal{A}}\times\boldsymbol{\mathcal{B}}\right)=\boldsymbol{\mathcal{B}}\cdot\left(\nabla\times\boldsymbol{\mathcal{A}}\right)-\boldsymbol{\mathcal{A}}\cdot\left(\nabla\times \boldsymbol{\mathcal{B}}\right)$, it follows that
\begin{equation}
\!\nabla\cdot \boldsymbol{\mathcal{F}}=\!-\!
\left[\boldsymbol{\mathcal{E}}\cdot\partial_t\left(\nabla\times\boldsymbol{\mathcal{D}}\right)+\boldsymbol{\mathcal{H}}\cdot\partial_t\left(\nabla\times\boldsymbol{\mathcal{B}}\right)+\mathcal{S}_{\rm \mathcal{J}}\right]\!/2,
\label{Eq4}
\end{equation}
where $\mathcal{S}_{\rm \mathcal{J}}=\boldsymbol{\mathcal{E}}\cdot\left(\nabla\times \boldsymbol{\mathcal{J}}\right)$ is the current-related source-like contribution. Taking into account the general structure of the continuity equation \cite{SM}, Eq. \eqref{Eq4} can be recast as
\begin{equation}
\nabla\cdot\boldsymbol{\mathcal{F}}+\partial_t\mathcal{C}=\mathcal{S},
\label{Eq5}
\end{equation}
where 
\begin{eqnarray}
\!\!\!\!\mathcal{C}&\equiv&\frac{1}{2}\left[\boldsymbol{\mathcal{E}}\cdot\left(\nabla\times \boldsymbol{\mathcal{D}}\right)+\boldsymbol{\mathcal{H}}\cdot\left(\nabla\times \boldsymbol{\mathcal{B}}\right)\right],
\label{Eq6}\\
\!\!\!\!\mathcal{S}&\equiv&\frac{1}{2}\left[\partial_t\boldsymbol{\mathcal{E}}\cdot\left(\nabla\times \boldsymbol{\mathcal{D}}\right)+\partial_t \boldsymbol{\mathcal{H}}\cdot\left(\nabla\times \boldsymbol{\mathcal{B}}\right)-\mathcal{S}_{\rm \mathcal{J}}\right],
\label{Eq7}
\end{eqnarray}
are the optical chirality density and the source-like terms, respectively. It is worth remarking that the above expressions represent the most general result for the conservation law of optical chirality, without any restrictions on the nature of the medium, i.e., they are valid regardless of the linearity, homogeneity, isotropy, or dispersion. However, they differ significantly from the previously established continuity equation \cite{Lipkin1964,Tang2010,Bliokh2011,Schaferling},
\begin{equation}
\nabla\cdot\boldsymbol{\mathcal{F}}+\partial_t\mathcal{C}_{\rm vacuum}=\mathcal{S}_{\rm vacuum},
\label{Eq8}
\end{equation}
where $\mathcal{C}_{\rm vacumm}$ is the optical chirality density as defined in Eq. \eqref{Eq1}, and $\mathcal{S}_{\rm vacuum}\equiv -\left[\boldsymbol{\mathcal{J}}\cdot\left(\nabla\times \boldsymbol{\mathcal{E}}\right)+\boldsymbol{\mathcal{E}}\cdot\left(\nabla\times\boldsymbol{\mathcal{J}}\right)\right]/2$ is the source-like term in free space. As discussed in Sec.$\,\,$II of the Supplemental Material \cite{SM}, the essential discrepancy arises on account of the dispersion-related terms. In particular, it is easy to prove that $\mathcal{C}=\mathcal{C}_{\rm vacuum}+\mathcal{C}_{\rm medium}$, where $\mathcal{C}_{\rm medium}\equiv\left[\boldsymbol{\mathcal{E}}\cdot\left(\nabla\times \boldsymbol{\mathcal{P}}\right)+\mu_0 \boldsymbol{\mathcal{H}}\cdot\left(\nabla\times \boldsymbol{\mathcal{M}}\right)\right]/2$, and  $\boldsymbol{\mathcal{P}}$ and $\boldsymbol{\mathcal{M}}$ are the macroscopic polarization and magnetization fields. Strikingly, up to our knowledge, these considerations have never been properly analyzed in previous approaches \cite{Tang2010,Bliokh2011,Philbin2013,Choi2012,Poulikakos2016}. In fact, even though both the dispersion-related and the dissipation terms are explicitly disregarded in Eq. \eqref{Eq8}, it has been widely used for investigating chirality and chiroptical effects in media where the permittivity is highly dispersive, including plasmonic nanostructures as well as metamaterials \cite{Schaferling2012,Meinzer2013,Valev2013,Nesterov2016,Collins2017,Luo2017,Hentschel2017,Schaferling}. Thus, as shown below, the dispersion of the material systems brings about important corrections into the original expressions for the optical chirality density [compare Eqs. \eqref{Eq1} and \eqref{Eq6}] and the source-like terms of the continuity equation [compare Eqs. \eqref{Eq5} and \eqref{Eq8}], and hence, it must be generally considered.

\textit{Optical chirality density in lossless dispersive media: Brillouin's approach.\textemdash}For monochromatic electric and magnetic fields in free space, $\boldsymbol{\mathcal{E}}({\bf r},t)=\text{Re}{\left[{\bf E}({\bf r})e^{-i\omega t}\right]}$ and $\boldsymbol{\mathcal{H}}({\bf r},t)=\text{Re}{\left[{\bf H}({\bf r})e^{-i\omega t}\right]}$, the time-averaged optical chirality density is given by \cite{Tang2010,Bliokh2011}
\begin{equation}
C_{\rm vacuum}=\frac{\omega}{2c^2}\text{Im}{\left[{\bf E}\cdot{\bf H}^*\right]},
\label{Eq9}
\end{equation}
where bold letters stand for complex field amplitudes and the asterisk denotes complex conjugation. A straightforward calculation allows us to show that, for freely propagating electromagnetic plane waves, the maximum value of $C$ is achieved for CPL:
\begin{equation}
C_{\;\rm vacuum}^{(\pm){\rm CPL}}= \pm\frac{\omega}{2c^2}\frac{1}{Z_0}\left|{\bf E}\right|^2,
\label{Eq10}
\end{equation}
where $Z_0\equiv\sqrt{\mu_0/\varepsilon_0}$ is the vacuum impedance, and the signs $+$ and $-$ correspond to left- and right-handed CPL.

In general, CPL is considered as the paradigmatic example of field displaying optical chirality, and has been widely used for chiroptical measurements \cite{Hendry2010,Tang2011}. Unfortunately, mainly due to the mismatch between the scales of the wavelength of light and the typical size of chiral objects \cite{Yang2009}, chiral responses are inherently very small \cite{Barron,Rhee2013}. To overcome this drawback, several efforts have been undertaken in the last decades for improving the detection schemes \cite{Tang2011,Kramer2017,Rosales2012,Corbaton2016}, with special emphasis on metallic nanostructures, which are regarded as well suited platforms for strengthening chiroptical light-matter interactions \cite{Schaferling2012,Meinzer2013,Valev2013,Nesterov2016,Collins2017,Luo2017,Hentschel2017,Schaferling}.

Metals are inherently absorptive and highly dispersive. Something similar happens with semiconductors at energies around the band gap. These features are characterized in terms of the electric permittivity $\varepsilon$ (and eventually with the magnetic permeability $\mu$) depending on the frequency $\omega$. According to the Kramer-Kronig relations \cite{Jackson}, dispersion is necessarily tied to dissipation. Thus, in order to avoid misleading outcomes, the analysis of the local dynamical properties have to be carefully carried out from a material standpoint as well. This is well known for the electromagnetic field energy in metals, for which a general treatment has been developed \cite{Jackson,Novotny}. Indeed, in a lossless dispersive medium the energy density is described by the Brillouin's formula \cite{Brillouin,Landau}. Following a similar procedure we may then obtain a closed expression for the optical chirality density. For simplicity, we will assume a linear, homogeneous and isotropic medium such that ${\bf D}=\varepsilon_0\varepsilon(\omega){\bf E}$ and ${\bf B}=\mu_0\mu(\omega){\bf H}$. From the continuity equation as given in Eq. \eqref{Eq4}, and using the Fourier transforms, the instantaneous distribution of the optical chirality density can be obtained by integrating 
$\boldsymbol{\mathcal{E}}\cdot\partial_t\left(\nabla\times\boldsymbol{\mathcal{D}}\right)$ and $\boldsymbol{\mathcal{H}}\cdot\partial_t\left(\nabla\times\boldsymbol{\mathcal{B}}\right)$ over time. It should be noted that the integral convergence is constrained by the \textit{slowly varying amplitude approximation} \cite{Novotny}. Within this assumption, the electric contribution reads as
\begin{equation}
\mathcal{C}^{\rm elec}\!=\!\frac{i}{2c^2}\!\iint{\!\left[\frac{\omega^2\varepsilon_{\omega}\mu_{\omega}}{\omega'+\omega}\right]\!{\bf E}_{\omega'}\!\cdot\!{\bf H}_{\omega} e^{-i(\omega'+\omega) t}d\omega'd\omega},
\label{Eq11}
\end{equation}
where the subscripts denote the frequency dependence. By proceeding in the same way for the magnetic contribution, summing up both expressions, and integrating them properly over the frequencies $\omega$ and $\omega'$ \cite{SM}, we can get the time-averaged optical chirality density in a lossless dispersive media:
\begin{equation}
C_{\rm lossless}=\text{Re}{\left[n(\omega)\tilde{n}(\omega)\right]}C_{\rm vacuum}=\frac{\omega}{2}\frac{\text{Im}{\left[{\bf E}\cdot{\bf H}^*\right]}}{v_p(\omega)v_g(\omega)},
\label{Clossless}
\end{equation}
where $v_p(\omega)\equiv c/\text{Re}{[n(\omega)]}$ and $v_g(\omega)\equiv c/\text{Re}{[\tilde{n}(\omega)]}$, are the phase and group velocities \cite{Boyd2009,Gerasik2010}, respectively, which are in turn expressed in terms of the complex-valued phase refractive index $n(\omega)\equiv\sqrt{\varepsilon(\omega)\mu(\omega)}$ and the corresponding dispersion-modified group index $\tilde{n}(\omega)\equiv n(\omega)+\omega\left[\partial n(\omega)/\partial \omega\right]$. A detailed description of the above derivation as well as the current-related contribution can be found in Sec.$\,\,$III.B of the Supplemental Material \cite{SM}. 

It should be noted that the same expression for the optical chirality density was previously obtained, but using a more complicated approach (see Eq. (33) in Ref. \cite{Philbin2013}). Importantly, this definition [Eq. \eqref{Clossless}] reduces to the standard result [Eq. \eqref{Eq9}] for freely propagating optical fields, i.e., when $n=1$. Furthermore, it is important to emphasize the dependence of Eq. \eqref{Clossless} on the dispersion-related phase and group velocities. From this simple relation, it is easy to realize that we may enhance the optical chirality in artificially engineered materials directly by lowering both velocities \cite{Boyd2009,Gerasik2010}. This is specifically accomplished in the vicinity of the resonance frequency, i.e., in the anomalous dispersion region [see upper panel of Fig. \ref{fig1}]. However, in a dispersive and lossy media, there are certain frequency ranges where the precise physical meaning of the group velocity turns out to be somewhat unclear \cite{Brillouin,Landau}, and then the result given by Eq. \eqref{Clossless} may not be valid. 

\begin{figure}[t!]
	\includegraphics[width=0.93\linewidth]{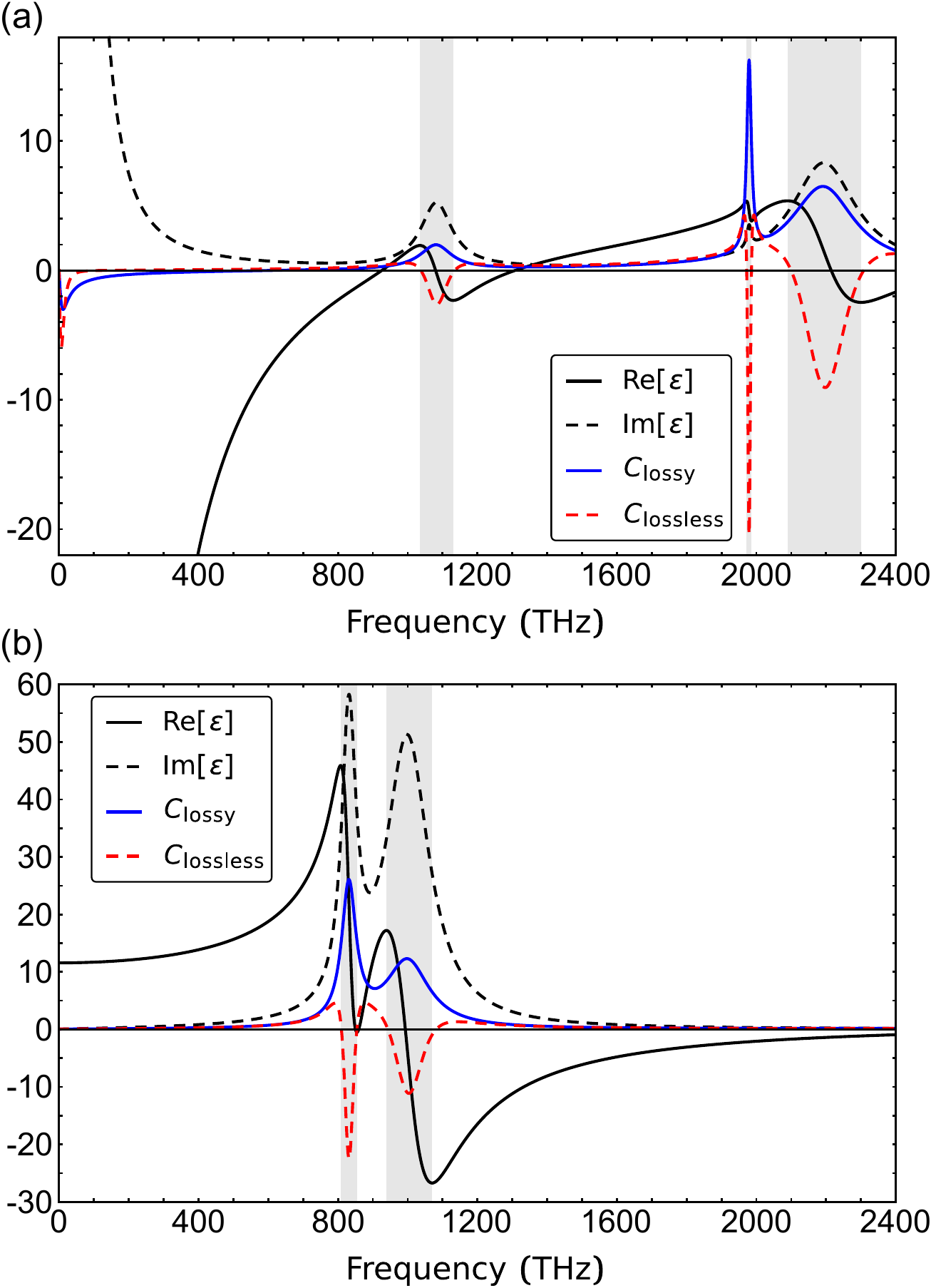} 
	\caption{Optical chirality density for (a) silver [Ag], and (b) silicon [Si]. Material parameters describing the permittivities are taken from Refs. \cite{Rakic1998} and \cite{Palik}, respectively. For comparison we represent the results for lossless (red dashed lines) [Eq. \eqref{Clossless}] and lossy (blue solid lines) [Eq. \eqref{Clossy}] dispersive media. The gray shaded areas indicate the spectral ranges with anomalous dispersion behavior.}
	\label{fig2}
\end{figure}
\textit{Optical chirality density in lossy dispersive media: Loudon's approach.\textemdash}A more physically realistic description of dispersive media requires careful considerations of dissipative effects. In this regard, as previously reported (see, e.g., Refs. \cite{Loudon1970,Ruppin2002}), the expression for the energy density in dispersive and lossy media crucially depend on the specific model characterizing the medium. This information is enclosed within the material parameters, $\varepsilon$ and $\mu$ \cite{Jackson,Novotny}.

In classical theory, $\varepsilon$ can be modeled as a collection of Lorentz oscillators \cite{Novotny,Maier}:
\begin{equation}
\varepsilon_{\rm Drude-Lorentz}(\omega)=1-\sum\limits_{n}{\frac{f_n\omega_{\rm p}^2}{\omega^2-\omega_n^2+i\omega\gamma_n}},
\label{Eq13}
\end{equation}
where $f_n$, $\omega_{\rm p}$, $\omega_n$, $\omega$ and $\gamma_n$ are, respectively, the relative strength of the oscillators, the plasma frequency, the $n$th resonance frequency, the excitation frequency, and the $n$th damping constant. This multi-resonant model has been proved to fit very well with the experimental data \cite{Sehmi2017,Rakic1998,Palik}, and thus, it can be regarded as a generic approach for characterizing the electric response of any material system for any frequency and bandwidth. A similar expression can also be introduced for the magnetic permeability $\mu$, e.g., when describing negative-index metamaterials consisting of arrays of split-ring resonators or fishnet-like structures \cite{Smith2000,GarciaMeca2011,Yoo2014}. It should be noted that the latter expression for $\varepsilon$ follows from the dynamic equation of the polarization field:
\begin{equation}
\frac{\partial^2\boldsymbol{\mathcal{P}}_n}{\partial t^2}+\gamma_n\frac{\partial \boldsymbol{\mathcal{P}}_n}{\partial t}+\omega_n^2\boldsymbol{\mathcal{P}}_n=\varepsilon_0f_n\omega_{\rm p}^2 \boldsymbol{\mathcal{E}}_{\rm loc},
\label{Eq14}
\end{equation}
with $\boldsymbol{\mathcal{E}}_{\rm loc}$ being the time-varying external electric field. This relation between the electric and the polarization field is actually the key point to get the general form of the energy density \cite{Loudon1970,Ruppin2002}. Likewise, taking into account the underlying mathematical structure of the continuity equation \cite{SM}, we can use Eq. \eqref{Eq14} (and the corresponding one for the magnetization field) to identify the electric (and the magnetic) contribution of the optical chirality density in dispersive and lossy media. To this aim, we start again from the continuity equation as given in Eq. \eqref{Eq4}. Attempting to find the total time derivative for the electric contribution, we have to express $\boldsymbol{\mathcal{E}}\cdot\partial_t\left(\nabla\times\boldsymbol{\mathcal{D}}\right)$ in terms of the electric and the polarization fields (and similarly for the magnetic contribution \cite{SM}):
\begin{equation}
\boldsymbol{\mathcal{E}}\cdot\partial_t\left(\nabla\times\boldsymbol{\mathcal{D}}\right)\!=\!\left[\varepsilon_0\boldsymbol{\mathcal{E}}\cdot\partial_t\!\left(\nabla\times \boldsymbol{\mathcal{E}}\right)+\boldsymbol{\mathcal{E}}\cdot\partial_t\!\left(\nabla\times \boldsymbol{\mathcal{P}}\right)\right].
\label{Eq15}
\end{equation}
In this way, we can also account for the influence of the medium on the chirality density. In the latter expression, the first term of the right-hand side can be rewritten as
\begin{equation}
\boldsymbol{\mathcal{E}}\cdot\partial_t \left(\nabla\times \boldsymbol{\mathcal{E}}\right) 
=\partial_t\left[\boldsymbol{\mathcal{E}}\cdot\left(\nabla\times \boldsymbol{\mathcal{E}}\right)\right]-\partial_t\boldsymbol{\mathcal{E}}\cdot\left(\nabla\times \boldsymbol{\mathcal{E}}\right),
\label{Eq16}
\end{equation}
thereby leading to a total time derivative plus a residual term. As shown at the end of Sec.$\,\,$III.A in the Supplemental Material \cite{SM}, this residual term exactly cancel with the one appearing for the magnetic contribution in vacuum, thus allowing us to recover the usual expression for the optical chirality in free space [Eq. \eqref{Eq1}]. On the other hand, the second term in the right-hand side of Eq. \eqref{Eq15} can be addressed by using the dynamic equation for the polarization field given in Eq. \eqref{Eq14} (see Sec.$\,\,$III.A in the Supplemental Material \cite{SM} for further details on this derivation). Following the same procedure for the magnetic contribution and summing up both expressions, we finally find that the time-averaged optical chirality density in a lossy dispersive medium is
\begin{equation}
C_{\rm lossy}\frac{\omega}{4c^2}\text{Im}{\left[\left(\varepsilon(\omega)\mu_{\rm eff}(\omega)+\varepsilon_{\rm eff}(\omega)\mu^*(\omega)\right){\bf E}\cdot{\bf H}^*\right]}.
\label{Clossy}
\end{equation}
where $\varepsilon_{\rm eff}$ and $\mu_{\rm eff}$ are the real-valued effective material parameters, which are defined as
\begin{subequations}
\begin{align}
&\varepsilon_{\rm eff}(\omega)\equiv 1+\sum\limits_{n}{\left(\chi'_n+2\omega\chi_n''/\gamma_n\right)},
\label{Eq18a}\\
&\mu_{\rm eff}(\omega)\equiv 1+\sum\limits_{n}{\left(\xi'_n+2\omega\xi_n''/\tilde{\gamma}_n\right)},
\label{Eq18b}
\end{align}
\end{subequations}
with $\chi=\chi'+i\chi''\equiv\varepsilon-1$ and $\xi=\xi'+i\xi''\equiv\mu-1$ being the electric and magnetic susceptibilities. Furthermore, as pointed out in the Supplemental Material, there are also a current-related contribution which should be included.

As shown in Fig. \ref{fig1}, both of the above approaches yield different results. Indeed, whereas $C_{\rm lossless}$ [Eq. \eqref{Clossless}] can display both positive and negative values, the total contribution of $C_{\rm lossy}$ [Eq. \eqref{Clossy}], remains always positive, with a minimum value of $C_{\rm vacuum}$ that is reached in the high-frequency limit. The largest discrepancies occur close to the resonance frequency. Still, the peaks for both approaches are almost equal in absolute value. These signatures can also be appreciated in Fig. \ref{fig2}, where we plot the optical chirality density of silver and silicon, as examples of metal and semiconductor, respectively. Both materials have been modeled using Eq. \eqref{Eq13} with parameters taken from Refs. \cite{Rakic1998} and \cite{Palik}. From the results shown in Fig. \ref{fig2} it is worth emphasizing that $C_{\rm lossless}$ overlaps almost exactly with $C_{\rm lossy}$ for all frequencies, except in the vicinity of the region of anomalous dispersion, i.e., where $d\varepsilon'/d\omega<0$. There, the curves turn out to be drastically separated from each other, thereby highlighting the importance of considering dissipative effects.

Equation \eqref{Clossy} is the main result of this work. To the best of our knowledge, it provides the most general definition for the optical chirality density in dispersive and lossy media, being applicable to any material system including plasmonic nanostructures \cite{Schaferling} and left-handed metamaterials \cite{Yoo2014}. Yet, our definition differs significantly from the standard formula for optical fields in free space \cite{Tang2010,Bliokh2011}, and even from previous suggestions attempting to tackle optical chirality in dispersive media \cite{Philbin2013,Poulikakos2016}. As discussed in Sec.$\,\,$III of the Supplemental Material \cite{SM}, the distinction between our result and those found in the aforementioned approaches essentially arises from considering properly the dynamic response of the time-dependent electromagnetic fields within a dispersive medium. In this regard, it should be noted that the time derivative of the fields $\boldsymbol{\mathcal{D}}$ and $\boldsymbol{\mathcal{B}}$, must be expressed as convolution integrals in the time domain. Furthermore, in this particular case, regarding lossy dispersive media, the mathematical structure of the continuity equation plays a central role in the identification of the optical chirality density as a conserved dynamical property. 

\textit{Summary.\textemdash}We have carried out a theoretical analysis of the conservation law for the optical chirality. Taking advantage from previous approaches addressing the electromagnetic energy density, we have also provided a parallel derivation for the optical chirality both in lossless and lossy dispersive media.  Remarkably, our description is completely general, i.e., is valid for arbitrarily varying radiation fields, and can be applied to any medium, including dielectrics, semiconductors, as well as highly lossy material systems such as metals and metamaterials with negative refractive index. In view of the growing interest on chirality and chiral light-matter interaction, we hope that these results will aid the development of plasmonic and metamaterial nanostructures for advanced chiroptical applications \cite{Barr2018,Yoo2014}, especially in the context of enhanced enantioselectivity, detection and characterization of chiral biomolecules via specifically designed chiral and nonchiral structures \cite{Etxarri2013,Pellegrini2017}.

\begin{acknowledgments}
The authors are grateful to C. Garc\'ia-Meca for valuable comments and discussions. This work was supported by fundings from Ministerio de Econom\'ia y Competitividad (MINECO) of Spain under Contract No. TEC2014-51902-C2-1-R.
\end{acknowledgments}

\clearpage
\appendix
\onecolumngrid
\renewcommand\theequation{S\arabic{equation}}
\setcounter{equation}{0}
\renewcommand\thepage{S.\arabic{page}}
\setcounter{page}{1}
\begin{center}
{\large \textbf{Supplemental Material:\\Optical Chirality in Dispersive and Lossy Media}}
\end{center}
\begin{center}
J. Enrique V\'azquez\,-Lozano$\blue{^*}$ and Alejandro Mart\'inez$\blue{^\dagger}$\\
{\small\textit{Nanophotonics Technology Center, Universitat Polit\`ecnica de Val\`encia, Camino de Vera s/n, 46022 Valencia, Spain}}\\
(Dated: \today)
\end{center}
\begin{quotation}
Building on the earlier approaches addressing the electromagnetic energy density in dispersive and lossy media, in this supplemental material we provide a parallel derivation for the optical chirality density considering dispersion as well as dissipation. To this aim, we first elaborate on the most general form of the continuity equation for the optical chirality. As a result we find a complete description, valid for arbitrary structured optical fields regardless of the time dependence. The only restriction we need to impose relies on the characterization of the material parameters, which must be fitted by Lorentzian line shapes. In the particular case of time-harmonic fields in a linear medium, we show that our findings are perfectly consistent with the ones so far established for electromagnetic fields in free space.
\end{quotation}
\section{I. ENERGY CONSERVATION IN DISPERSIVE AND LOSSY MEDIA: ENERGY DENSITY AND FLOW}
The energy conservation law, also referred to as the Poynting's theorem, is a fundamental statement of electrodynamics \cite{Poynting1884sm}. Even though at a first glance it seems to be a somewhat trivial concern, the situation may become more complicated when considering an optical field traveling through a dispersive medium, i.e., a material system whose electromagnetic response depends on the frequency of the applied field. In the time domain field representation this means that the material response is not instantaneous, but it relies on the electromagnetic fields at all previous time instants. Indeed, according to the basic properties of Fourier transform, this is described in terms of a convolution in the time domain of material equations \cite{Jacksonsm}. This behavior is actually characterized by means of the electric permittivity $\varepsilon$, and the magnetic permeability, $\mu$, both depending on the frequency $\omega$, and described by the corresponding constitutive relations, which account for the influence of the electromagnetic radiation on matter. Importantly, it should be noted that, strictly speaking, dispersion is necessarily tied to dissipation. This connection is well established by the so-called Kramer-Kronig relations \cite{Gorkunov2015sm}, according to which the real and imaginary parts of the material parameters ($\varepsilon(\omega)=\varepsilon'+i\varepsilon''$ and $\mu(\omega)=\mu'+i\mu''$) appear to be coupled together. In addition, it has been demonstrated that Kramer-Kronig relations underpin the fundamental principle of causality, and hence, initial assumptions regarding dispersion and dissipation have to be carefully considered, otherwise they may lead to misleading outcomes. Still, we can find many cases where is assumed a dispersive medium neglecting the losses.

The relatively recent advent of left-handed materials \cite{Veselago1968sm,Pendry2000sm,Lezec2007sm,Smith2000sm}, and the development of metamaterial photonics in general \cite{Engheta2007sm,Valentine2009sm,GarciaMeca2011sm,Sun2012sm}, have led to reexamine both the treatment and the formulation \cite{Loudon1970sm,Ruppin2002sm,Cui2004sm,Boardman2006sm,Luan2009sm,Raman2010sm,Webb2010sm,Rosa2010sm,Nunes2011sm,Shin2012sm,Tretyakov2005sm} of the field energy in dispersive media \cite{Poynting1884sm,Jacksonsm,Brillouinsm,Landausm}. In this respect, one of the main difficulties arises on account of a \textit{bona fide} interpretation about each of the contributions appearing in the continuity equation, e.g., the sense of negative values for the energy density (in regards to the thermodynamic significance of the stability conditions \cite{Landausm}) in the anomalous dispersion region (near the resonance frequencies) \cite{Askne1970sm,Huang2001sm,Ziolkowski2001sm,Shelby2001sm}, or the attainable superluminal (or even backward) light propagation \cite{Wang2000sm,Feigenbaum2009sm,Glasgow2001sm,Dolling2006sm}. Furthermore, owing to the close relation between the Poynting vector and the electromagnetic momentum \cite{Nelson1991sm}, the long-standing controversy known as the Abraham-Minkowski dilemma \cite{Barnet2010asm}, that still remains as a subject of current interest around the proper definition of the linear momentum for optical field in media, may also be an additional source of puzzling discussions and conclusions \cite{Barnett2010bsm,Bliokh2017sm,Silveirinha2017sm}.

Below, we briefly sketch the derivation outlined by Loudon \cite{Loudon1970sm} (and extended by Ruppin \cite{Ruppin2002sm}) in order to settle out an appropriate framework for the subsequent analysis dealing with the optical chirality conservation law. Moreover, aiming to find a closed expression, we also look into the adiabatic-like classical approaches put forward by Brillouin \cite{Brillouinsm}, and Landau \cite{Landausm}. For the sake of clarity, throughout this work we will use a distinct notation either for the real-valued electromagnetic radiation in the time domain, or for the complex fields in the frequency domain. It is well known that both representations are related to each other via (inverse) Fourier transform \cite{Novotnysm}:
\begin{equation}
\boldsymbol{\mathcal{F}}({\bf r},t)=\int_{-\infty}^{+\infty}{{\bf F}({\bf r},\omega)e^{-i\omega t}d\omega}\qquad \Longleftrightarrow \qquad {\bf F}({\bf r},\omega)=\frac{1}{2\pi}\int_{-\infty}^{+\infty}{\boldsymbol{\mathcal{F}}({\bf r},t)e^{i\omega t}dt},
\label{S1}
\end{equation}
where the vector field $\boldsymbol{\mathcal{F}}({\bf r},t)$ (or ${\bf F}({\bf r},\omega)$) stands for the electric field $\boldsymbol{\mathcal{E}}$, the electric displacement $\boldsymbol{\mathcal{D}}$, the magnetic field $\boldsymbol{\mathcal{H}}$, or the magnetic induction $\boldsymbol{\mathcal{B}}$ (or their corresponding complex-like counterparts). Moreover, so as to avoid cumbersome notations, we shall omit the argument in the field expressions. 

Taking into account the above convention, the general expression for the time-dependent energy conservation law can be straightforwardly stemmed from the vector identity $\nabla\cdot\left(\boldsymbol{\mathcal{A}}\times\boldsymbol{\mathcal{B}}\right)=\boldsymbol{\mathcal{B}}\cdot\left(\nabla\times\boldsymbol{\mathcal{A}}\right)-\boldsymbol{\mathcal{A}}\cdot\left(\nabla\times \boldsymbol{\mathcal{B}}\right)$, with the aid of the structural (or curl-like) Maxwell's equations  $\nabla\times \boldsymbol{\mathcal{E}}=-\partial_t \boldsymbol{\mathcal{B}}$ and $\nabla\times \boldsymbol{\mathcal{H}}=\partial_t \boldsymbol{\mathcal{D}}+\boldsymbol{\mathcal{J}}$:
\begin{equation}
\nabla\cdot \boldsymbol{\mathcal{S}}=-\boldsymbol{\mathcal{E}}\cdot \partial_t \boldsymbol{\mathcal{D}}-\boldsymbol{\mathcal{H}}\cdot \partial_t\boldsymbol{\mathcal{B}}-\boldsymbol{\mathcal{J}}\cdot \boldsymbol{\mathcal{E}},
\label{S2}
\end{equation}
where $\boldsymbol{\mathcal{S}}\equiv \boldsymbol{\mathcal{E}}\times \boldsymbol{\mathcal{H}}$ is the Poynting vector, which represent the energy flux density \cite{Poynting1884sm}, and $\boldsymbol{\mathcal{J}}$ is the electric current density. The previous equation is known as the differential form of Poynting's theorem and is an absolutely general result, i.e., holds for both lossy and lossless media \cite{Poynting1884sm,Jacksonsm,Landausm,Novotnysm}. Besides to express the energy balance over the whole system, due to its mathematical structure,
\begin{equation}
\nabla\cdot \text{[FLUX]}+\partial_t\text{[CONSERVED PROPERTY]}=\text{[SOURCE or SINK]},
\label{S3}
\end{equation}
the continuity equation, in conjunction with Noether's theorem, constitutes a powerful tool for identifying and analyzing dynamical properties that are often, but not always, conserved, such as the energy, the linear momentum, or the optical chirality, among many others. Therefore, in the simplest case of linear and nondispersive media, i.e, if $\boldsymbol{\mathcal{D}}=\varepsilon_0\varepsilon \boldsymbol{\mathcal{E}}$ and $\boldsymbol{\mathcal{B}}=\mu_0\mu \boldsymbol{\mathcal{H}}$, with $\varepsilon,\mu\in\mathbb{R}$, Eq. \eqref{S2} can be recast as
\begin{equation}
\nabla\cdot \boldsymbol{\mathcal{S}}+\partial_t\mathcal{W}=-\boldsymbol{\mathcal{J}}\cdot \boldsymbol{\mathcal{E}},
\label{S4}
\end{equation}
where $\mathcal{W}=\left[\boldsymbol{\mathcal{E}}\cdot \boldsymbol{\mathcal{D}}+\boldsymbol{\mathcal{H}}\cdot \boldsymbol{\mathcal{B}}\right]/2$ is the instantaneous energy density stored (or conserved) in the electromagnetic field, and $\boldsymbol{\mathcal{J}}\cdot \boldsymbol{\mathcal{E}}$ represent the power loss dissipated to the medium on account of external current sources \cite{Jacksonsm,Landausm,Novotnysm}. However, a more physically realistic description requires careful considerations of dispersion and, consequently, dissipation \cite{footnote1sm}. In such a case, as pointed out above, the dynamic response of a time-dependent electric field passing through a dispersive medium is expressed as \cite{Jacksonsm}
\begin{equation}
\boldsymbol{\mathcal{D}}({\bf r},t)=\int_{-\infty}^{t}{\varepsilon(t-t')\boldsymbol{\mathcal{E}}({\bf r},t')dt'}=\int_{-\infty}^{+\infty}{\varepsilon(t-t')\boldsymbol{\mathcal{E}}({\bf r},t')dt'}.
\label{S5}
\end{equation}
Notice that a similar expression can be posed for the magnetic contribution in terms of $\mu$. Here it is worth observing that Eq. \eqref{S2} involves the time derivative of the fields $\boldsymbol{\mathcal{D}}$ and $\boldsymbol{\mathcal{B}}$, which are in turn written as the convolution integral given in Eq. \eqref{S5}. This is in fact the main difficulty for obtaining, in this case, a general and closed expression for the instantaneous electromagnetic energy density. Nevertheless, there are several routes to overcome the evaluation of the time derivative \cite{Brillouinsm,Landausm,Loudon1970sm,Ruppin2002sm,Shin2012sm,Cui2004sm,Boardman2006sm,Luan2009sm,Webb2010sm,Nunes2011sm,Tretyakov2005sm,Raman2010sm,Rosa2010sm}, each of them subject to well distinct prescriptions relying upon assumptions concerning both the characteristics of the medium, and the time dependence of the electromagnetic fields. In the present work we pursue an approach as general as possible, so that it can be applied to an arbitrarily varying radiation field which propagates within a truly dispersive media, namely, including the absorption losses as well. However it should be emphasized that the expressions accounting for both the energy density stored and the dissipation, will crucially depend upon the specific features of the model characterizing the medium \cite{Tretyakov2005sm,Boardman2006sm,Luan2009sm}. Hence, for practical purposes, we will focus on the pragmatic treatment provided by Loudon \cite{Loudon1970sm}, which consider an absorbing classical dielectric with a single resonance frequency, i.e., a Lorentz-like medium. This approach has been further extended by many other authors to account for dispersive magnetic permeabilities \cite{Ruppin2002sm,Cui2004sm,Tretyakov2005sm,Boardman2006sm,Luan2009sm}, as well as the possibility of multiple resonance frequencies describing the influence of interband transition effects \cite{Oughstun1988sm,Raman2010sm,Nunes2012sm,Shin2012sm}. For completeness, and for convenience in subsequent analysis, we shall take into account these generalizations throughout this work.

Optical properties of metals (often regarded as the paradigmatic example of dispersive media) may be classically modeled by distinguishing between the response owing to the free electrons moving within the conduction band (usually referred to as the intraband effects), and that of the bound carriers, giving rise to the interband transitions \cite{Maiersm,Novotnysm}. This can be generally described as a collection of $N$ oscillators in the following form \cite{Rakic1998sm}:
\begin{equation}
\varepsilon_{\rm Drude-Lorentz}(\omega)=1-\frac{f_0\omega_{\rm p}^2}{\omega^2+i\omega\gamma_0}-\sum\limits_{n=1}^{N-1}{\frac{f_n\omega_{\rm p}^2}{\omega^2-\omega_n^2+i \omega\gamma_n}}=1-\sum\limits_{n=0}^{N-1}{\frac{f_n\omega_{\rm p}^2}{\omega^2-\omega_n^2+i\omega\gamma_n}},
\label{S6}
\end{equation}
where $f_n$, $\omega_{\rm p}$, $\omega_n$, $\omega$ and $\gamma_n$ are, respectively, the relative strength of the oscillators, the plasma frequency, the $n$th resonance (or restoring) frequency, the excitation frequency, and the $n$th damping constant (or characteristic collision frequency). Note that, since the Drude model only holds for intraband effects, there is no resonant behavior for the first term in Eq. \eqref{S6}, and thus $\omega_0\equiv0$. In any case, each of the Lorentz-like oscillators is described as a pole in the dielectric function [Eq. \eqref{S6}], and can be readily achieved from the equation of motion for a bound electron with undamped resonance frequency $\omega_n$, experiencing a damping force characterized by $\gamma_n$, and subjected to a time-varying external electric field $\boldsymbol{\mathcal{E}}_{\rm loc}$ \cite{footnote2sm}:
\begin{equation}
\frac{\partial^2{\bf r}_n}{\partial t^2}+\gamma_n\frac{\partial {\bf r}_n}{\partial t}+\omega_n^2{\bf r}_n=-\frac{q_{\rm e}}{m_{\rm e}} \boldsymbol{\mathcal{E}}_{\rm loc},
\label{S7}
\end{equation}
where $m_{\rm e}$, $q_{\rm e}$ and ${\bf r}_n$ stand for the effective mass, charge, and the displacement of the electrons, respectively. If we further interpret each mode as an electron gas of uniform density $\rho_n$, the collective effect emerging from all individual displacement leads to a polarization field $\boldsymbol{\mathcal{P}}_n=\rho_n{\bf p}_n=\left(-q_{\rm e}\rho_n\right){\bf r}_n$, where ${\bf p}_n$ is the electric dipole moment associated to the $n$th mode. Hence, Eq. \eqref{S7} can be rewritten as
\begin{equation}
\frac{\partial^2\boldsymbol{\mathcal{P}}_n}{\partial t^2}+\gamma_n\frac{\partial \boldsymbol{\mathcal{P}}_n}{\partial t}+\omega_n^2\boldsymbol{\mathcal{P}}_n=\varepsilon_0f_n\omega_{\rm p}^2 \boldsymbol{\mathcal{E}}_{\rm loc},
\label{S8}
\end{equation}
where $\varepsilon_0$ is the electric permittivity of vacuum,  $\omega_{\rm p}\equiv\sqrt{q_{\rm e}^2\rho_{\rm e}/(m_{\rm e}\varepsilon_0)}$, and $f_n\equiv\rho_n/\rho_{\rm e}$, with $\rho_{\rm e}$ being the total density of Drude and Lorentz oscillators. By solving Eq. \eqref{S8} for fields with harmonic time dependence of the form $ e^{-i\omega t}$, one can obtain the individual contribution of the oscillators:
\begin{equation}
{\bf P}_n=\rho_n\alpha_n^{\rm (e)}(\omega){\bf E}_{\rm loc}=\varepsilon_0\left[\frac{f_n\omega_{\rm p}^2}{\omega_n^2-\omega^2-i\omega\gamma_n}\right] {\bf E}_{\rm loc},
\label{S9}
\end{equation}
where $\alpha_n^{\rm (e)}(\omega)$ is the electric dipole polarizability. Therefore, each of the Lorentz poles contribute to the electric permittivity given in Eq. \eqref{S6} in such a way that
\begin{equation}
{\bf D}=\varepsilon_0\left[1-\sum_{n=0}^{N-1}{\frac{f_n\omega_{\rm p}^2}{\omega^2-\omega_n^2+i\omega\gamma_n}}\right]{\bf E},
\label{S10}
\end{equation}
where the total polarization field is defined as ${\bf P}\equiv\sum_n{\braket{{\bf P}_n}}$, and the macroscopic electric field is ${\bf E}\equiv\braket{{\bf E}_{\rm loc}}$, with the angular brackets indicating an average over space \cite{footnote2sm}. This procedure is specific for getting the electric permittivity in a linear medium modeled as a combination of Drude and Lorentz oscillators \cite{Rakic1998sm,Nunes2012sm,Oughstun1988sm,Maiersm}. Still, provided the magnetic permeability can be properly tailored by Lorentzian line shapes, it may also be extended to magnetically dispersive media \cite{Ruppin2002sm,Cui2004sm,Tretyakov2005sm,Boardman2006sm,Luan2009sm} by means of the corresponding dynamic equation associated to the magnetization field $\boldsymbol{\mathcal{M}}_n$: 
\begin{equation}
\frac{\partial^2\boldsymbol{\mathcal{M}}_n}{\partial t^2}+\tilde{\gamma}_n\frac{\partial \boldsymbol{\mathcal{M}}_n}{\partial t}+\tilde{\omega}_n^2\boldsymbol{\mathcal{M}}_n=\tilde{f}_n\tilde{\omega}_n^2 \boldsymbol{\mathcal{H}}_{\rm loc},
\label{S11}
\end{equation}
where, analogously to the previous case, $\tilde{\omega}_n$, $\tilde{\gamma}_n$, and $\tilde{f}_n$ are, respectively, the $n$th resonance frequency of the magnetic dipole oscillators, the $n$th magnetic damping constant and the magnetic-like oscillator strength. It should be noted that in earlier works addressing magnetic dispersion (e.g., when characterizing split-ring resonator-based structures) only a single Lorentzian resonance takes place \cite{Ruppin2002sm,Cui2004sm,Tretyakov2005sm,Boardman2006sm,Luan2009sm}. Nonetheless, for completeness, we shall express the magnetic permeability by considering the possible appearance of $N$ magnetic dipole oscillators of the form
\begin{equation}
{\bf M}_n=\left[\frac{\tilde{f}_n\tilde{\omega}^2_n}{\tilde{\omega}_n^2-\omega^2-i\omega\tilde{\gamma}_n}\right] {\bf H}_{\rm loc}.
\label{S12}
\end{equation}
Hence, if we define the total magnetization field as ${\bf M}\equiv\sum_n{\braket{{\bf M}_n}}$, and the macroscopic magnetic field ${\bf H}\equiv\braket{{\bf H}_{\rm loc}}$ \cite{footnote2sm}, the magnetic permeability is given by
\begin{equation}
\mu_{\rm Drude-Lorentz}(\omega)=1-\sum\limits_{n=0}^{N-1}{\frac{\tilde{f}_n\tilde{\omega}_n^2}{\omega^2-\tilde{\omega}_n^2+i\omega\tilde{\gamma}_n}}.
\label{S13}
\end{equation}

After this brief outline, we are now in the position to show the expression for the energy conservation law in dispersive and lossy media, fully accounting for both the energy density and the power density lost due to light-matter interactions. Whatever the specific electromagnetic properties of the medium, it can always be represented in terms of the polarization and the magnetization vector fields as follows:
\begin{eqnarray}
\boldsymbol{\mathcal{D}}&=&\varepsilon_0\boldsymbol{\mathcal{E}}+\boldsymbol{\mathcal{P}},
\label{S14}\\
\boldsymbol{\mathcal{B}}&=&\mu_0\boldsymbol{\mathcal{H}}+\mu_0\boldsymbol{\mathcal{M}}.
\label{S15}
\end{eqnarray}
Then, the first two terms in the right-hand side of Eq. \eqref{S2} can be straightforwardly evaluated by using the dynamic equations for the polarization and magnetization fields [Eqs. \eqref{S8} and \eqref{S11}], with the aid of Eqs. \eqref{S14} and \eqref{S15}:
\begin{eqnarray}
\boldsymbol{\mathcal{E}}\cdot\frac{\partial \boldsymbol{\mathcal{D}}}{\partial t}&=&\partial_t\left\{\frac{\varepsilon_0}{2}\boldsymbol{\mathcal{E}}^2+\sum_{n=0}^{N-1}{\frac{1}{2\varepsilon_0f_n\omega_{\rm p}^2}}\left[\left(\partial_t\braket{\boldsymbol{\mathcal{P}}_n}\right)^2+\omega_n^2\braket{\boldsymbol{\mathcal{P}}_n}^2\right]\right\}+\sum_{n=0}^{N-1}{\frac{\gamma_n}{\varepsilon_0f_n\omega_{\rm p}^2}\left(\partial_t \braket{\boldsymbol{\mathcal{P}}_n}\right)^2},
\label{S16}\\
\boldsymbol{\mathcal{H}}\cdot\frac{\partial \boldsymbol{\mathcal{B}}}{\partial t}&=&\partial_t\left\{\frac{\mu_0}{2}\boldsymbol{\mathcal{H}}^2+\sum_{n=0}^{N-1}{\frac{\mu_0}{2\tilde{f}_n\tilde{\omega_n^2}}\left[\left(\partial_t \braket{\boldsymbol{\mathcal{M}}_n}\right)^2+\tilde{\omega}_n^2\left(\braket{\boldsymbol{\mathcal{M}}_n}\right)^2\right]}\right\}+\sum_{n=0}^{N-1}{\frac{\mu_0\tilde{\gamma}_n}{\tilde{f}_n\tilde{\omega_n^2}}\left(\partial_t\braket{\boldsymbol{\mathcal{M}}_n}\right)^2}.
\label{S17}
\end{eqnarray}
The above results are completely general, and they are applicable independently of the time dependence considered for the electromagnetic fields. The only restriction we need to impose is that the material parameters of the medium must be fitted by Lorentzian line shapes. In such a case, they allow us to identify accurately both the electric and the magnetic contributions to the energy density stored either by the fields themselves or by the material system, as well as the terms accounting for the power loss densities:
\begin{eqnarray}
\boldsymbol{\mathcal{E}}\cdot\partial_t\boldsymbol{\mathcal{D}}&=&\partial_t\left[\mathcal{W}_{\rm vacuum}^{\rm electric}+\mathcal{W}_{\rm medium}^{\rm electric}\right]+\mathcal{L}^{\rm electric}_{\rm energy},
\label{S18}\\
\boldsymbol{\mathcal{H}}\cdot\partial_t\boldsymbol{\mathcal{B}}&=&\partial_t\left[\mathcal{W}_{\rm vacuum}^{\rm magnetic}+\mathcal{W}_{\rm medium}^{\rm magnetic}\right]+\mathcal{L}^{\rm magnetic}_{\rm energy}.
\label{S19}
\end{eqnarray}
Taking into account the above expressions, it is worth mentioning that both the energy and losses can be directly identified by comparing them with the general structure of the continuity equation given in Eq. \eqref{S3}:
\begin{equation}
\nabla\cdot \boldsymbol{\mathcal{S}}+\partial_t\left[\mathcal{W}_{\rm vacuum}^{\rm electric}+\mathcal{W}_{\rm medium}^{\rm electric}+\mathcal{W}_{\rm vacuum}^{\rm magnetic}+\mathcal{W}_{\rm medium}^{\rm magnetic}\right]=-\left[\mathcal{L}^{\rm magnetic}_{\rm energy}+\mathcal{L}^{\rm magnetic}_{\rm energy}+\mathcal{L}^{\rm current}_{\rm energy}\right],
\label{S20}
\end{equation}
where $\mathcal{L}^{\rm current}_{\rm energy}=\boldsymbol{\mathcal{J}}\cdot \boldsymbol{\mathcal{E}}$. Notice that in the case of only one Lorentz oscillator, i.e., $N=1$, the results reduce to those of Ref. \cite{Ruppin2002sm}. A deeper understanding of the role played by these terms deserves further efforts beyond the scope of this work (c.f. Refs. \cite{Loudon1970sm,Ruppin2002sm,Cui2004sm,Tretyakov2005sm,Boardman2006sm,Luan2009sm,Nunes2012sm,Oughstun1988sm,Raman2010sm,Shin2012sm}). Despite that, turns out to be convenient to compare the results provided by this procedure, wherein the dispersion and dissipation are truly considered, with those stemmed from the classical approaches put forward by Brillouin and Landau \cite{Landausm,Brillouinsm}. To this aim we calculate the time average of the energy density by considering time harmonic fields. Thus, from Eqs. \eqref{S16} and \eqref{S17}, we can obtain
\begin{eqnarray}
\braket{\mathcal{W}^{\rm electric}}_{\rm T}&=&\frac{\varepsilon_0}{4}\left|{\bf E}\right|^2\left[1+\sum\limits_{n=0}^{N-1}{\frac{f_n\left(\omega^2+\omega_n^2\right)\omega_{\rm p}^2}{\left(\omega_n^2-\omega^2\right)^2+\omega^2\gamma_n^2}}\right]=\frac{\varepsilon_0\varepsilon_{\rm eff}(\omega)}{4}\left|{\bf E}\right|^2, \label{S21}\\
\braket{\mathcal{W}^{\rm magnetic}}_{\rm T}&=&\frac{\mu_0}{4}\left|{\bf H}\right|^2\left[1+\sum\limits_{n=0}^{N-1}{\frac{\tilde{f}_n\left(\omega^2+\tilde{\omega}_n^2\right)\tilde{\omega}_n^2}{\left(\tilde{\omega}_n^2-\omega^2\right)^2+\omega^2\tilde{\gamma}_n^2}}\right]=\frac{\mu_0\mu_{\rm eff}(\omega)}{4}\left|{\bf H}\right|^2,
\label{S22}
\end{eqnarray}
where the angle brackets indicate time averaging over one period of oscillation, and $\varepsilon_{\rm eff}$ and $\mu_{\rm eff}$ are the real-valued effective material parameters (i.e., the electric permittivity and magnetic permeability), which are defined as
\begin{equation}
\varepsilon_{\rm eff}(\omega)\equiv 1+\sum\limits_{n=0}^{N-1}{\left(\chi'_n+\frac{2\omega\chi_n''}{\gamma_n}\right)},\qquad \qquad
\mu_{\rm eff}(\omega)\equiv 1+\sum\limits_{n=0}^{N-1}{\left(\xi'_n+\frac{2\omega\xi_n''}{\tilde{\gamma}_n}\right)},
\label{S23}
\end{equation}
with $\chi=\sum_n{\chi_n}=\chi'+i\chi''\equiv\varepsilon-1$ and $\xi=\sum_n{\xi_n}=\xi'+i\xi''\equiv\mu-1$ being the electric and magnetic susceptibilities:
\begin{eqnarray}
\chi_n&=&\frac{f_n\omega_{\rm p}^2\left(\omega_n^2-\omega^2\right)}{\left(\omega_n^2-\omega^2\right)^2+\omega^2\gamma_n^2}+i\left(\frac{f_n\omega_{\rm p}^2\omega\gamma_n}{\left(\omega_n^2-\omega^2\right)^2+\omega^2\gamma_n^2}\right),
\label{S24}\\
\xi_n&=&\frac{\tilde{f}_n\tilde{\omega}_n^2\left(\tilde{\omega}_n^2-\omega^2\right)}{\left(\tilde{\omega}_n^2-\omega^2\right)^2+\omega^2\tilde{\gamma}_n^2}+i\left(\frac{\tilde{f}_n\tilde{\omega}_n^2\omega\tilde{\gamma}_n}{\left(\tilde{\omega}_n^2-\omega^2\right)^2+\omega^2\tilde{\gamma}_n^2}\right).
\label{S25}
\end{eqnarray}
On the other hand, according to the approach of Brillouin, it is known that the time average of the energy density is 
\begin{equation}
\braket{\mathcal{W}_{\rm Brillouin}^{\rm electric}}=\frac{\varepsilon_0\tilde{\varepsilon}(\omega)}{4}\left|{\bf E}\right|^2,\qquad \qquad 
\braket{\mathcal{W}_{\rm Brillouin}^{\rm magnetic}}=\frac{\mu_0\tilde{\mu}(\omega)}{4}\left|{\bf H}\right|^2.
\label{S26}
\end{equation}
where $\tilde{\varepsilon}(\omega)\equiv\varepsilon'+\omega[\partial\varepsilon'/\partial\omega]$ and $\tilde{\mu}(\omega)\equiv\mu'+\omega[\partial\mu'/\partial\omega]$ are the dispersion-modified material parameters. After some calculations it is tedious but straightforward to show that $\tilde{\varepsilon}(\omega)=\varepsilon_{\rm eff}(\omega)$ and $\tilde{\mu}(\omega)=\mu_{\rm eff}(\omega)$ if and only if 
\begin{equation}
\left(\omega^2+\omega_n^2\right)\gamma_n^2=0,\qquad \text{and} \qquad \left(\omega^2+\tilde{\omega}_n^2\right)\tilde{\gamma}_n^2=0,
\label{S27}
\end{equation}
namely, if $\gamma_n=0$ and/or $\omega=\pm i\omega_n$, and $\tilde{\gamma}_n=0$ and/or $\omega=\pm i\tilde{\omega}_n$, for all $n$. Hence, both approaches exactly coincide only in the case of lossless media, and/or otherwise under the a priori physically meaningless condition of imaginary operating frequencies \cite{Nunes2011sm}.
\section{II. CONTINUITY EQUATION FOR OPTICAL CHIRALITY IN DISPERSIVE AND LOSSY MEDIA}
Building on the above scheme, below we will provide a step-by-step derivation of the correct form of the continuity equation (or conservation law) for the optical chirality in dispersive and lossy media. As was previously pointed out, in a proper sense, dispersion and dissipation are needly related to each other, and therefore we are forced to consider them together. Furthermore, regarding the corresponding dynamical property, the source-like contributions, containing loss or gain terms, afford valuable insights into fundamental aspects of light-matter interaction \cite{Nienhuis2016sm,Corbaton2017sm}.  

The literature concerning with optical chirality and its interaction with matter mostly deals with electromagnetic fields in free space \cite{Lipkin1964sm,Tang2010sm,Tang2011sm,Bliokh2011sm,Coles2012sm}. That is why we often find several expressions for the chirality flux density where the fields $\boldsymbol{\mathcal{B}}$ and $\boldsymbol{\mathcal{H}}$ are used interchangeably yielding the same result, as of course it must. However, this does not occur in dispersive media, and special care should be taken in dealing with electromagnetic fields either in free-space ($\boldsymbol{\mathcal{E}}$ and $\boldsymbol{\mathcal{H}}$), or within a material system ($\boldsymbol{\mathcal{D}}$ and $\boldsymbol{\mathcal{B}}$). For symmetry reasons, we may heuristically assume that the chirality flux density reads as
\begin{equation}
\boldsymbol{\mathcal{F}}\equiv\frac{1}{2}\left[\boldsymbol{\mathcal{E}}\times\left(\nabla\times \boldsymbol{\mathcal{H}}\right)-\boldsymbol{\mathcal{H}}\times\left(\nabla\times \boldsymbol{\mathcal{E}}\right)\right].
\label{S28}
\end{equation}
This axiomatic definition is in fact that used in Ref. \cite{Poulikakos2016sm}, and coincides with that originally introduced by Tang and Cohen \cite{Tang2010sm} for electromagnetic fields in free space. Following a similar procedure as before for the energy conservation law, from Eq. \eqref{S28} we can readily calculate the divergence of the chirality flux density as:
\begin{eqnarray}
\nonumber\nabla\cdot \boldsymbol{\mathcal{F}}&=&\frac{1}{2}\left[\boldsymbol{\mathcal{H}}\cdot\nabla\times\left(\nabla\times \boldsymbol{\mathcal{E}}\right)-\boldsymbol{\mathcal{E}}\cdot\nabla\times\left(\nabla\times \boldsymbol{\mathcal{H}}\right)\right]\\
&=&-\frac{1}{2}\left[\boldsymbol{\mathcal{H}}\cdot\partial_t\left(\nabla\times\boldsymbol{\mathcal{B}}\right)+\boldsymbol{\mathcal{E}}\cdot\partial_t\left(\nabla\times\boldsymbol{\mathcal{D}}\right)+\boldsymbol{\mathcal{E}}\cdot\left(\nabla\times \boldsymbol{\mathcal{J}}\right)\right].
\label{S29}
\end{eqnarray}
For obtaining an expression with the structure of the continuity equation as given in Eq. \eqref{S3}, we need to find a total time derivative term. Thus, from the latter expression it is straightforward to show that
\begin{eqnarray}
\partial_t\left[\boldsymbol{\mathcal{H}}\cdot\left(\nabla\times \boldsymbol{\mathcal{B}}\right)\right]&=&\boldsymbol{\mathcal{H}}\cdot\partial_t\left(\nabla\times \boldsymbol{\mathcal{B}}\right)+\partial_t \boldsymbol{\mathcal{H}}\cdot\left(\nabla\times\boldsymbol{\mathcal{B}}\right),
\label{S30}\\
\partial_t\left[\boldsymbol{\mathcal{E}}\cdot\left(\nabla\times \boldsymbol{\mathcal{D}}\right)\right]&=&\boldsymbol{\mathcal{E}}\cdot\partial_t\left(\nabla\times \boldsymbol{\mathcal{D}}\right)+\partial_t \boldsymbol{\mathcal{E}}\cdot\left(\nabla\times\boldsymbol{\mathcal{D}}\right).
\label{S31}
\end{eqnarray}
By means of the above expressions we can recast Eq. \eqref{S29} as
\begin{equation}
\nabla\cdot\boldsymbol{\mathcal{F}}+\partial_t\mathcal{C}=\mathcal{S},
\label{S32}
\end{equation}
where the optical chirality density and the source-like terms have been defined as
\begin{eqnarray}
\mathcal{C}&\equiv&\frac{1}{2}\left[\boldsymbol{\mathcal{E}}\cdot\left(\nabla\times \boldsymbol{\mathcal{D}}\right)+\boldsymbol{\mathcal{H}}\cdot\left(\nabla\times \boldsymbol{\mathcal{B}}\right)\right],
\label{S33}\\
\mathcal{S}&\equiv&\frac{1}{2}\left[\partial_t\boldsymbol{\mathcal{E}}\cdot\left(\nabla\times \boldsymbol{\mathcal{D}}\right)+\partial_t \boldsymbol{\mathcal{H}}\cdot\left(\nabla\times \boldsymbol{\mathcal{B}}\right)-\boldsymbol{\mathcal{E}}\cdot\left(\nabla\times \boldsymbol{\mathcal{J}}\right)\right].
\label{S34}
\end{eqnarray}

It is worth pointing out that the above expressions represent the most general result for the conservation law of optical chirality without restrictions on the nature of the medium, namely, Eqs. \eqref{S32}, \eqref{S33} and \eqref{S34} are valid regardless of the linearity, homogeneity, isotropy, or dispersion. Nonetheless, they differ considerably from the previously established continuity equation \cite{Lipkin1964sm,Tang2010sm,Tang2011sm,Bliokh2011sm,Coles2012sm,Philbin2013sm,Poulikakos2016sm,Cameron2017sm}:
\begin{equation}
\nabla\cdot\boldsymbol{\mathcal{F}}_{\rm vacuum}+\partial_t\mathcal{C}_{\rm vacuum}=\mathcal{S}_{\rm vacuum},
\label{S35}
\end{equation}
where $\boldsymbol{\mathcal{F}}_{\rm vacuum}\equiv\boldsymbol{\mathcal{F}}$, $\mathcal{C}_{\rm vacuum}\equiv\left[\varepsilon_0\boldsymbol{\mathcal{E}}\cdot\left(\nabla\times \boldsymbol{\mathcal{E}}\right)+\mu_0\boldsymbol{\mathcal{H}}\cdot\left(\nabla\times \boldsymbol{\mathcal{H}}\right)\right]/2$, and $\mathcal{S}_{\rm vacuum}\equiv -\left[\boldsymbol{\mathcal{J}}\cdot\left(\nabla\times \boldsymbol{\mathcal{E}}\right)+\boldsymbol{\mathcal{E}}\cdot\left(\nabla\times\boldsymbol{\mathcal{J}}\right)\right]/2$ are, respectively, the chirality flux density, the optical chirality density and the source-like terms in free space. Curiously, albeit Eq. \eqref{S35} was initially posed for optical fields in vacuum, has been widely used for investigating chiroptical effects occurring in material systems, including metals as well. However, it can be demonstrated that the general result given in Eq. \eqref{S32} reduces to Eq. \eqref{S35} only for linear and nonabsorbing media, i.e., as long as $\boldsymbol{\mathcal{D}}=\varepsilon_0\varepsilon \boldsymbol{\mathcal{E}}$ and $\boldsymbol{\mathcal{B}}=\mu_0\mu \boldsymbol{\mathcal{H}}$, with $\varepsilon=\mu=1$:
\begin{eqnarray}
\nonumber\mathcal{C}&=&\frac{1}{2}\left[\boldsymbol{\mathcal{E}}\cdot\left(\nabla\times \boldsymbol{\mathcal{D}}\right)+\boldsymbol{\mathcal{H}}\cdot\left(\nabla\times \boldsymbol{\mathcal{B}}\right)\right]\\
&=&\frac{1}{2}\left[\varepsilon_0 \boldsymbol{\mathcal{E}}\cdot\left(\nabla\times \boldsymbol{\mathcal{E}}\right)+\mu_0 \boldsymbol{\mathcal{H}}\cdot\left(\nabla\times \boldsymbol{\mathcal{H}}\right)\right]=\mathcal{C}_{\rm vacuum},
\label{S36}
\end{eqnarray}
and
\begin{eqnarray}
\nonumber\mathcal{S}&=&\frac{1}{2}\left[\partial_t\boldsymbol{\mathcal{E}}\cdot\left(\nabla\times \boldsymbol{\mathcal{D}}\right)+\partial_t \boldsymbol{\mathcal{H}}\cdot\left(\nabla\times \boldsymbol{\mathcal{B}}\right)-\boldsymbol{\mathcal{E}}\cdot\left(\nabla\times \boldsymbol{\mathcal{J}}\right)\right]\\
\nonumber&=&\frac{1}{2}\left[\varepsilon_0\partial_t\boldsymbol{\mathcal{E}}\cdot\left(-\mu_0\partial_t\boldsymbol{\mathcal{H}}\right)+\mu_0\partial_t \boldsymbol{\mathcal{H}}\cdot\left(\varepsilon_0\partial_t \boldsymbol{\mathcal{E}}+\boldsymbol{\mathcal{J}}\right)-\boldsymbol{\mathcal{E}}\cdot\left(\nabla\times \boldsymbol{\mathcal{J}}\right)\right]\\
\nonumber&=&\frac{1}{2}\left[\mu_0\partial_t \boldsymbol{\mathcal{H}}\cdot \boldsymbol{\mathcal{J}}-\boldsymbol{\mathcal{E}}\cdot\left(\nabla\times \boldsymbol{\mathcal{J}}\right)\right]\\
&=&-\frac{1}{2}\left[\boldsymbol{\mathcal{J}}\cdot\left(\nabla\times \boldsymbol{\mathcal{E}}\right)+\boldsymbol{\mathcal{E}}\cdot\left(\nabla\times \boldsymbol{\mathcal{J}}\right)\right]=\mathcal{S}_{\rm vacuum},
\label{S37}
\end{eqnarray}
where $\mu_0\partial_t\boldsymbol{\mathcal{H}}=-\left(\nabla\times \boldsymbol{\mathcal{E}}\right)$. Hence, we stress here the relevance for considering dispersive-like terms. This becomes much more evident by rewriting the above expressions in terms of those for vacuum. Indeed, since
\begin{eqnarray}
\boldsymbol{\mathcal{E}}\cdot\left(\nabla\times \boldsymbol{\mathcal{D}}\right)&=&\boldsymbol{\mathcal{D}}\cdot\left(\nabla\times \boldsymbol{\mathcal{E}}\right)+\nabla\cdot\left(\boldsymbol{\mathcal{P}}\times \boldsymbol{\mathcal{E}}\right),
\label{S38}\\
\boldsymbol{\mathcal{H}}\cdot\left(\nabla\times \boldsymbol{\mathcal{B}}\right)&=&\boldsymbol{\mathcal{B}}\cdot\left(\nabla\times \boldsymbol{\mathcal{H}}\right)+\mu_0\nabla\cdot\left(\boldsymbol{\mathcal{M}}\times \boldsymbol{\mathcal{H}}\right),
\label{S39}
\end{eqnarray}
the optical chirality density given in Eq. \eqref{S33} can be generally expressed as
\begin{equation}
\mathcal{C}=\frac{1}{2}\left\{\boldsymbol{\mathcal{D}}\cdot\left(\nabla\times \boldsymbol{\mathcal{E}}\right)+\boldsymbol{\mathcal{B}}\cdot\left(\nabla\times \boldsymbol{\mathcal{H}}\right)+\nabla\cdot\left[\boldsymbol{\mathcal{P}}\times \boldsymbol{\mathcal{E}}+\mu_0\left(\boldsymbol{\mathcal{M}}\times \boldsymbol{\mathcal{H}}\right)\right]\right\}=\mathcal{C}_{\rm vacuum}+\mathcal{C}_{\rm medium},
\label{S40}
\end{equation}
where $\mathcal{C}_{\rm medium}\equiv\left[\boldsymbol{\mathcal{E}}\cdot\left(\nabla\times \boldsymbol{\mathcal{P}}\right)+\mu_0 \boldsymbol{\mathcal{H}}\cdot\left(\nabla\times \boldsymbol{\mathcal{M}}\right)\right]/2$. On the other hand, the source-like term can also be recast as
\begin{equation}
\mathcal{S}=\mathcal{S}_{\rm vacuum}+\mathcal{S}_{\rm medium},
\label{S41}
\end{equation}
where $\mathcal{S}_{\rm medium}\equiv\left\{\partial_t\boldsymbol{\mathcal{E}}\cdot\left(\nabla\times \boldsymbol{\mathcal{P}}\right) -\partial_t \boldsymbol{\mathcal{P}}\cdot\left(\nabla\times \boldsymbol{\mathcal{E}}\right)+\mu_0\left[\partial_t \boldsymbol{\mathcal{H}}\cdot\left(\nabla\times\boldsymbol{\mathcal{M}}\right)-\partial_t \boldsymbol{\mathcal{M}}\cdot\left(\nabla\times \boldsymbol{\mathcal{H}}\right)\right]\right\}/2$. Therefore, by assuming a monochromatic optical field represented by ${\bf E}(\omega)=\left[{\bf E}\delta(\omega-\omega_0)+{\bf E}^*\delta(\omega+\omega_0)\right]/2$, it can be demonstrated that in a linear medium
\begin{eqnarray}
\braket{\boldsymbol{\mathcal{E}}\cdot\left(\nabla\times \boldsymbol{\mathcal{P}}\right)}_{\rm T}&=&\frac{\varepsilon_0\mu_0}{2}\text{Im}{\left[\omega\left(1-\varepsilon(\omega)\right)\mu(\omega){\bf E}^*\cdot{\bf H}\right]}+\frac{i\varepsilon_0\left({\bf E}\times{\bf E}^*\right)}{2}\text{Im}{\left[\nabla\varepsilon(\omega)\right]}, 
\label{S42}\\
\mu_0\braket{\boldsymbol{\mathcal{H}}\cdot\left(\nabla\times \boldsymbol{\mathcal{M}}\right)}_{\rm T} &=&\frac{\varepsilon_0\mu_0}{2}\text{Im}{\left[\omega\left(1-2\varepsilon^*(\omega)\right)\left(\mu^*(\omega)-1\right){\bf E}^*\cdot{\bf H}\right]}+\frac{i\mu_0\left({\bf H}\times{\bf H}^*\right)}{2}\text{Im}{\left[\nabla\mu(\omega)\right]},
\label{S43}
\end{eqnarray}
thus showing that
\begin{equation}
\braket{\mathcal{C}_{\rm medium}}_{\rm T}=\frac{1}{4c^2}\text{Im}{\left\{\left[\frac{i\sigma^*\xi^*}{\varepsilon_0}-\omega\chi\mu-\omega\varepsilon^*\xi^*\right]{\bf E}^*\cdot{\bf H}\right\}}+\frac{i\varepsilon_0\left({\bf E}\times{\bf E}^*\right)}{4}\text{Im}{\left[\nabla\varepsilon\right]}+\frac{i\mu_0\left({\bf H}\times{\bf H}^*\right)}{4}\text{Im}{\left[\nabla\mu\right]},
\label{S44}
\end{equation}
where $\sigma(\omega)=i\varepsilon_0\omega\left[1-\varepsilon(\omega)\right]$ is the complex-valued conductivity. Likewise, the time average over the terms comprising the source-like contributions yields
\begin{eqnarray}
\braket{\partial_t\boldsymbol{\mathcal{E}}\cdot\left(\nabla\times \boldsymbol{\mathcal{P}}\right)}_{\rm T}&=&-\frac{\varepsilon_0\mu_0}{2}\text{Re}{\left[\omega^2\chi(\omega)\mu(\omega){\bf E}^*\cdot{\bf H}\right]}+\frac{i\omega\varepsilon_0\left({\bf E}\times{\bf E}^*\right)}{2}\text{Re}{\left[\nabla\varepsilon(\omega)\right]},
\label{S45}\\
\braket{\partial_t\boldsymbol{\mathcal{P}}\cdot\left(\nabla\times \boldsymbol{\mathcal{E}}\right)}_{\rm T}&=&-\frac{\varepsilon_0\mu_0}{2}\text{Re}{\left[\omega^2\chi^*(\omega)\mu(\omega){\bf E}^*\cdot{\bf H}\right]},
\label{S46}\\
\braket{\mu_0\partial_t\boldsymbol{\mathcal{H}}\cdot\left(\nabla\times \boldsymbol{\mathcal{M}}\right)}_{\rm T}&=&\frac{\varepsilon_0\mu_0}{2}\text{Re}{\left[\omega^2\left(2\varepsilon^*(\omega)-1\right)\xi^*(\omega){\bf E}^*\cdot{\bf H}\right]}+\frac{i\omega\mu_0\left({\bf H}\times{\bf H}^*\right)}{2}\text{Re}{\left[\nabla\mu(\omega)\right]},
\label{S47}\\
\braket{\mu_0\partial_t\boldsymbol{\mathcal{M}}\cdot\left(\nabla\times \boldsymbol{\mathcal{H}}\right)}_{\rm T}&=&\frac{\varepsilon_0\mu_0}{2}\text{Re}{\left[\omega^2\left(2\varepsilon^*(\omega)-1\right)\xi(\omega){\bf E}^*\cdot{\bf H}\right]},
\label{S48}
\end{eqnarray}
and then
\begin{eqnarray}
\braket{\partial_t\boldsymbol{\mathcal{E}}\cdot\left(\nabla\times \boldsymbol{\mathcal{P}}\right)-\partial_t\boldsymbol{\mathcal{P}}\cdot\left(\nabla\times \boldsymbol{\mathcal{E}}\right)}_{\rm T}&=&\varepsilon_0\mu_0\omega^2\text{Im}{\left[\varepsilon\right]}\text{Im}{\left[\mu{\bf E}^*\cdot{\bf H}\right]}+\frac{i\omega\varepsilon_0\left({\bf E}\times{\bf E}^*\right)}{2}\text{Re}{\left[\nabla\varepsilon\right]},
\label{S49}\\
\mu_0\braket{\partial_t\boldsymbol{\mathcal{H}}\cdot\left(\nabla\times \boldsymbol{\mathcal{M}}\right)-\partial_t\boldsymbol{\mathcal{M}}\cdot\left(\nabla\times \boldsymbol{\mathcal{H}}\right)}_{\rm T}&=&\varepsilon_0\mu_0\omega^2\text{Im}{\left[\mu\right]}\text{Im}{\left[\left(2\varepsilon^*-1\right){\bf E}^*\cdot{\bf H}\right]}+\frac{i\omega\mu_0\left({\bf H}\times{\bf H}^*\right)}{2}\text{Re}{\left[\nabla\mu\right]}.
\label{S50}
\end{eqnarray}
For the sake of simplicity, let us focus our attention here in the case of homogeneous media, i.e., $\varepsilon\neq\varepsilon({\bf r})$, and $\mu\neq\mu({\bf r})$, in such a way that $\nabla\varepsilon=\nabla\mu=0$. In this simple case it can be found that
\begin{equation}
\braket{\mathcal{C}_{\rm medium}}_{\rm T}=\frac{1}{4c^2}\text{Im}{\left\{\left[\frac{i\sigma^*\xi^*}{\varepsilon_0}-\omega\chi\mu-\omega\varepsilon^*\xi^*\right]{\bf E}^*\cdot{\bf H}\right\}},
\label{S51}
\end{equation}
and
\begin{equation}
\braket{\mathcal{S}_{\rm medium}}_{\rm T}=\frac{\omega^2}{2c^2}\left\{\text{Im}{\left[\varepsilon\right]}\text{Im}{\left[\mu{\bf E}^*\cdot{\bf H}\right]}+\text{Im}{\left[\mu\right]}\text{Im}{\left[\left(2\varepsilon^*-1\right){\bf E}^*\cdot{\bf H}\right]}\right\}.
\label{S52}
\end{equation}
\noindent By analyzing the latter results it can be shown that, irrespectively of the input polarization, $\braket{\mathcal{C}_{\rm medium}}_{\rm T}=0$ if $\varepsilon=\mu=1$, i.e., in vacuum. On the other hand, as long as the medium is nondispersive, i.e., provided that $\text{Im}{\left[\varepsilon\right]}=\text{Im}{\left[\mu\right]}=0$, $\braket{\mathcal{S}_{\rm medium}}_{\rm T}=0$.

Summarizing, in this section we have shown that the conservation law for the optical chirality established up to now (see, e.g., Refs. \cite{Lipkin1964sm,Tang2010sm,Tang2011sm,Bliokh2011sm,Coles2012sm,Philbin2013sm,Poulikakos2016sm,Cameron2017sm}) is only valid for electromagnetic waves in free space. Starting from the definition of the chirality flux density as given in Eq. \eqref{S28}, we have provided a detailed derivation showing that there are terms which have been neglected in previous works. Significantly, the dispersion of the material system leads to important corrections into the expressions for the optical chirality density as well as for the source-like terms of the continuity equation, which must be generally considered.
\section{III. OPTICAL CHIRALITY DENSITY IN LINEAR DISPERSIVE AND LOSSY MEDIA}
Following a similar approach as in previous studies \cite{Loudon1970sm,Ruppin2002sm,Cui2004sm,Tretyakov2005sm,Boardman2006sm,Luan2009sm,Raman2010sm,Shin2012sm,Rosa2010sm,Webb2010sm,Nunes2011sm}, we will properly derive the optical chirality density in dispersive and lossy media. As mentioned above for the case of energy, our results will be valid provided the material parameters can be modeled by Lorentzian line shapes. Within this approach, the expression for the chirality density does not rely on the time dependence of the electromagnetic fields. For comparison, we will also determine the optical chirality density via the classical procedure leading to the Brillouin formula for the energy density \cite{Brillouinsm,Landausm}. Just as previously found, there are certain conditions where both approaches exactly coincide. We will show that these conditions for the chirality density are analogous to that obtained for the energy density. 
\subsection{A. Optical chirality density in dispersive and lossy media: Loudon's approach}
Starting from the continuity equation as given in Eq. \eqref{S29} and attempting to express the electric contribution in terms of the electric and polarization fields, $\boldsymbol{\mathcal{E}}$ and $\boldsymbol{\mathcal{P}}$ respectively, it follows that 
\begin{equation}
\frac{1}{2}\boldsymbol{\mathcal{E}}\cdot\partial_t\left(\nabla\times\boldsymbol{\mathcal{D}}\right)=\frac{1}{2}\left[\varepsilon_0\boldsymbol{\mathcal{E}}\cdot\partial_t \left(\nabla\times \boldsymbol{\mathcal{E}}\right)+\boldsymbol{\mathcal{E}}\cdot\partial_t \left(\nabla\times \boldsymbol{\mathcal{P}}\right)\right].
\label{S53}
\end{equation}
In this expression the first term of the right-hand side can be rewritten as
\begin{eqnarray}
\frac{\varepsilon_0}{2}\boldsymbol{\mathcal{E}}\cdot\partial_t \left(\nabla\times \boldsymbol{\mathcal{E}}\right) 
&=&\frac{\varepsilon_0}{2}\left\{\partial_t\left[\boldsymbol{\mathcal{E}}\cdot\left(\nabla\times \boldsymbol{\mathcal{E}}\right)\right]-\partial_t\boldsymbol{\mathcal{E}}\cdot\left(\nabla\times \boldsymbol{\mathcal{E}}\right)\right\},
\label{S54}
\end{eqnarray}
thereby leading to a total time derivative term plus a residual one. As will be shown below, this residual term exactly cancel with the one appearing for the magnetic contribution in vacuum, thus allowing us to recover the usual expression for the optical chirality in free space. Likewise, the second term in the right-hand side of Eq. \eqref{S53} can be addressed by using the dynamic equation for the polarization field given in Eq. \eqref{S8}:
\begin{eqnarray}
\nonumber\frac{1}{2}\boldsymbol{\mathcal{E}}\cdot\partial_t \left(\nabla\times \boldsymbol{\mathcal{P}}\right)&=&\partial_t\left\{\sum_{n=0}^{N-1}{\frac{1}{2\varepsilon_0f_n\omega_{\rm p}^2}\left[\partial_t\braket{\boldsymbol{\mathcal{P}}_n}\cdot\left(\nabla\times \partial_t\braket{\boldsymbol{\mathcal{P}}_n}\right)+\omega_n^2\braket{\boldsymbol{\mathcal{P}}_n}\cdot\left(\nabla\times \braket{\boldsymbol{\mathcal{P}}_n}\right)\right]}\right\}\\
\nonumber&&-\sum_{n=0}^{N-1}{\frac{1}{2\varepsilon_0f_n\omega_{\rm p}^2}\left\{\omega_n^2\partial_t \braket{\boldsymbol{\mathcal{P}}_n}\cdot\left(\nabla\times \braket{\boldsymbol{\mathcal{P}}_n}\right)+\partial_t \braket{\boldsymbol{\mathcal{P}}_n}\cdot\left(\nabla\times\partial_t^2 \braket{\boldsymbol{\mathcal{P}}_n}\right)\right\}}\\
&&+\sum_{n=0}^{N-1}{\frac{1}{2\varepsilon_0f_n\omega_{\rm p}^2}\left\{\gamma_n\partial_t\braket{\boldsymbol{\mathcal{P}}_n}\cdot\left(\nabla\times \partial_t\braket{\boldsymbol{\mathcal{P}}_n}\right)\right\}}.
\label{S55}
\end{eqnarray}
Analogously, we can write the magnetic contribution in terms of $\boldsymbol{\mathcal{H}}$ and $\boldsymbol{\mathcal{M}}$:
\begin{equation}
\frac{1}{2}\boldsymbol{\mathcal{H}}\cdot\partial_t\left(\nabla\times\boldsymbol{\mathcal{B}}\right)=\frac{\mu_0}{2}\left[\boldsymbol{\mathcal{H}}\cdot\partial_t \left(\nabla\times \boldsymbol{\mathcal{H}}\right)+\boldsymbol{\mathcal{H}}\cdot\partial_t \left(\nabla\times\boldsymbol{\mathcal{M}}\right)\right],
\label{S56}
\end{equation}
where
\begin{eqnarray}
\!\!\!\!\!\!\!\!\!\!\!\!\frac{\mu_0}{2}\boldsymbol{\mathcal{H}}\cdot\partial_t \left(\nabla\times \boldsymbol{\mathcal{H}}\right) 
&=&\frac{\mu_0}{2}\left\{\partial_t\left[\boldsymbol{\mathcal{H}}\cdot\left(\nabla\times \boldsymbol{\mathcal{H}}\right)\right]-\partial_t\boldsymbol{\mathcal{H}}\cdot\left(\nabla\times \boldsymbol{\mathcal{H}}\right)\right\},
\label{S57}\\
\!\!\!\!\!\!\!\!\!\!\!\!\nonumber\frac{\mu_0}{2}\boldsymbol{\mathcal{H}}\cdot\partial_t\left(\nabla\times\boldsymbol{\mathcal{M}}\right)&=&\mu_0\partial_t\left\{\sum_{n=0}^{N-1}{\frac{1}{2\tilde{f}_n\tilde{\omega}_n^2}\left[\partial_t\braket{\boldsymbol{\mathcal{M}}_n}\cdot\left(\nabla\times \partial_t\braket{\boldsymbol{\mathcal{M}}_n}\right)+\tilde{\omega}_n^2\braket{\boldsymbol{\mathcal{M}}_n}\cdot\left(\nabla\times \braket{\boldsymbol{\mathcal{M}}_n}\right)\right]}\right\}\\
\nonumber&&-\sum_{n=0}^{N-1}{\frac{\mu_0}{2\tilde{f}_n\tilde{\omega}_n^2}{\left\{\tilde{\omega}_n^2\partial_t\braket{\boldsymbol{\mathcal{M}}_n}\cdot\left(\nabla\times\braket{\boldsymbol{\mathcal{M}}_n}\right)+\partial_t \braket{\boldsymbol{\mathcal{M}}_n}\cdot\left(\nabla\times\partial_t^2\braket{\boldsymbol{\mathcal{M}}_n}\right)\right\}}}\\
&&+\sum_{n=0}^{N-1}{\frac{\mu_0}{2\tilde{f}_n\tilde{\omega}_n^2}{\left\{2\tilde{\gamma}_n\partial_t \braket{\boldsymbol{\mathcal{M}}_n}\cdot\left(\nabla\times \partial_t \braket{\boldsymbol{\mathcal{M}}_n}\right)\right\}}}.
\label{S58}
\end{eqnarray}
Then, taking into account the structure of the continuity equation as given in Eq. \eqref{S3} we can easily identify the electric and magnetic contributions of the optical chirality density stored either by the fields or by the medium, as well as the terms accounting for the loss rate of the chirality density \cite{Nienhuis2016sm,Corbaton2017sm}:
\begin{eqnarray}
\frac{1}{2}\boldsymbol{\mathcal{E}}\cdot\partial_t\left(\nabla\times\boldsymbol{\mathcal{D}}\right)&=&\partial_t\left[\mathcal{C}_{\rm vacuum}^{\rm electric}+\mathcal{C}_{\rm medium}^{\rm electric}\right]+\mathcal{L}_{\rm chirality}^{\rm electric},
\label{S59}\\
\frac{1}{2}\boldsymbol{\mathcal{H}}\cdot\partial_t\left(\nabla\times\boldsymbol{\mathcal{B}}\right)&=&\partial_t\left[\mathcal{C}_{\rm vacuum}^{\rm magnetic}+\mathcal{C}_{\rm medium}^{\rm magnetic}\right]+\mathcal{L}_{\rm chirality}^{\rm magnetic}.
\label{S60}
\end{eqnarray}
From the above equations one can directly observe that $\boldsymbol{\mathcal{E}}\cdot\partial_t\left(\nabla\times\boldsymbol{\mathcal{D}}\right)$ and $\boldsymbol{\mathcal{H}}\cdot\partial_t\left(\nabla\times\boldsymbol{\mathcal{B}}\right)$ enclose generally the joint action of stored together with the dissipative (or gained) contributions. Aiming to compare these results with those obtained via the Fourier transform, we should calculate the time average of the optical chirality density by considering time harmonic fields in a linear medium. Therefore, from Eqs. \eqref{S54}, \eqref{S55}, \eqref{S57} and \eqref{S58}, and with the help of Eqs. \eqref{S9} and \eqref{S12}, it can be demonstrated that
\begin{equation}
\braket{\mathcal{C}^{\rm electric}_{\rm vacuum+medium}}_{\rm T}=\frac{\omega}{4c^2}\text{Im}{\left[\mu^*(\omega){\bf E}\cdot{\bf H}^*\right]}\varepsilon_{\rm eff}(\omega),
\label{S61}
\end{equation}
and
\begin{equation}
\braket{\mathcal{C}^{\rm magnetic}_{\rm vacuum+medium}}_{\rm T}=\frac{\omega}{4c^2}\text{Im}{\left[\varepsilon(\omega){\bf E}\cdot{\bf H}^*\right]}\mu_{\rm eff}(\omega),
\label{S62}
\end{equation}
where $\varepsilon_{\rm eff}$ and $\mu_{\rm eff}$ are the real-valued effective material parameters defined in Eq. \eqref{S23}. Hence, the sum of the above quantities allow us to write the total optical chirality density for dispersive and lossy media:
\begin{equation}
\braket{\mathcal{C}^{\rm electric+magnetic}}_{\rm T}=\frac{\omega}{4c^2}\text{Im}{\left[\left(\varepsilon(\omega)\mu_{\rm eff}(\omega)+\varepsilon_{\rm eff}(\omega)\mu^*(\omega)\right){\bf E}\cdot{\bf H}^*\right]}.
\label{S63}
\end{equation}

It should also be noticed that in the magnetic contribution there appears an additional term associated with the current density ${\bf J}$:
\begin{equation}
\braket{\mathcal{C}^{\rm current}_{\rm medium}}_{\rm T}=\frac{\omega}{4c^2}\text{Im}{\left[(\varepsilon(\omega)-1){\bf E}\cdot{\bf H}^*\right]}\mu_{\rm eff}(\omega).
\label{S64}
\end{equation}
Then, the complete expression for the optical chirality conservation law actually reads as follows:
\begin{equation}
\nabla\cdot \boldsymbol{\mathcal{F}}+\partial_t\left[\mathcal{C}^{\rm electric}_{\rm vacuum+medium}+\mathcal{C}^{\rm magnetic}_{\rm vacuum+medium}+\mathcal{C}^{\rm current}_{\rm medium}\right]=-\left[\mathcal{L}_{\rm chirality}^{\rm electric}+\mathcal{L}_{\rm chirality}^{\rm magnetic}+\mathcal{L}_{\rm chirality}^{\rm current}\right],
\label{S65}
\end{equation}
where $\mathcal{L}_{\rm chirality}^{\rm current}=\boldsymbol{\mathcal{E}}\cdot\left(\nabla\times \boldsymbol{\mathcal{J}}\right)/2$. However it is interesting to realize that $\mathcal{L}_{\rm chirality}^{\rm current}$ actually leads to conserved-like contributions. This can be easily shown by turning to the Drude model \cite{Maiersm}, where the current density satisfies
\begin{equation}
\frac{\partial\boldsymbol{\mathcal{J}}}{\partial t}+\gamma_0\boldsymbol{\mathcal{J}}=\varepsilon_0\omega_{\rm p}^2 \boldsymbol{\mathcal{E}}.
\end{equation}
Indeed, by using the latter dynamic equation, it follows that
\begin{equation}
\frac{1}{2}\boldsymbol{\mathcal{E}}\cdot\left(\nabla\times \boldsymbol{\mathcal{J}}\right)=\frac{1}{2\varepsilon_0\omega_{\rm p}^2}\left\{\partial_t\left[\boldsymbol{\mathcal{J}}\cdot\left(\nabla\times \boldsymbol{\mathcal{J}}\right)\right]-\boldsymbol{\mathcal{J}}\cdot\left(\nabla\times \partial_t \boldsymbol{\mathcal{J}}\right)+\gamma_0\boldsymbol{\mathcal{J}}\cdot\left(\nabla\times \boldsymbol{\mathcal{J}}\right)\right\}.
\label{S67}
\end{equation}
The source-like contribution given in Eq. \eqref{S67} may also be written as the field contributions [Eqs. \eqref{S59} and \eqref{S60}]:
\begin{equation}
\frac{1}{2}\boldsymbol{\mathcal{E}}\cdot\left(\nabla\times \boldsymbol{\mathcal{J}}\right)=\partial_t\left[\mathcal{C}^{\rm current}_{\rm sources}\right]+\mathcal{L}_{\rm chirality}^{\rm current}.
\label{S68}
\end{equation}
Taking into account the latter results, in a linear, homogeneous and isotropic medium it follows that
\begin{equation}
\braket{\mathcal{C}^{\rm current}_{\rm sources}}_{\rm T}=\frac{\omega\mu_0|\sigma_{\rm Drude}|^2}{4\varepsilon_0\omega_{\rm p}^2}\text{Im}{\left[\mu^*(\omega){\bf E}\cdot{\bf H}^*\right]}=\frac{\omega}{4c^2}\left[\frac{\omega_{\rm p}^2}{\omega^2+\gamma_0^2}\right]\text{Im}{\left[\mu^*(\omega){\bf E}\cdot{\bf H}^*\right]},
\label{S69}
\end{equation}
where it has been used that
\begin{equation}
{\bf J}=\left[\frac{i\omega\varepsilon_0\omega_{\rm p}^2}{\omega^2+i\omega\gamma_0}\right]{\bf E}=i\omega\varepsilon_0\left(1-\varepsilon_{\rm Drude}(\omega)\right){\bf E}\equiv \sigma_{\rm Drude}{\bf E}.
\label{S70}
\end{equation}
For simplicity we focus here on the simpler case of nonmagnetic medium, $\mu=\mu_{\rm eff}=1$, thus obtaining that
\begin{equation}
\braket{\mathcal{C}^{\rm current}_{\rm medium}}_{\rm T}+\braket{\mathcal{C}^{\rm current}_{\rm sources}}_{\rm T}=\frac{1}{4c^2}\left[\frac{\gamma_0\omega_{\rm p}^2}{\omega^2+\gamma_0^2}\right]\text{Re}{\left[{\bf E}\cdot{\bf H}^*\right]},
\label{S71}
\end{equation}
and
\begin{equation}
\braket{\mathcal{C}^{\rm electric}_{\rm vacuum+medium}}_{\rm T}+\braket{\mathcal{C}^{\rm magnetic}_{\rm vacuum+medium}}_{\rm T}=\frac{\omega}{2c^2}\text{Im}{\left[{\bf E}\cdot{\bf H}^*\right]}+\frac{1}{4c^2}\left[\frac{\gamma_0\omega_{\rm p}^2}{\omega^2+\gamma_0^2}\right]\text{Re}{\left[{\bf E}\cdot{\bf H}^*\right]}.
\label{S72}
\end{equation}
Notice that, since we are considering the Drude model, we should use Eqs. \eqref{S23}, \eqref{S24} and \eqref{S25} in the simplest case of $N=1$, $f_0=1$, and $\omega_n=\tilde{\omega}_n=0$ for all $n$. Summing up the above expressions we get the chirality density:
\begin{equation}
\braket{\mathcal{C}^{\rm electric}_{\rm vacuum+medium}}_{\rm T}+\braket{\mathcal{C}^{\rm magnetic}_{\rm vacuum+medium}}_{\rm T}+\braket{\mathcal{C}^{\rm current}_{\rm sources+medium}}_{\rm T}=\frac{\omega}{2c^2}\text{Im}{\left[{\bf E}\cdot{\bf H}^*\right]}+\frac{1}{2c^2}\left[\frac{\gamma_0\omega_{\rm p}^2}{\omega^2+\gamma_0^2}\right]\text{Re}{\left[{\bf E}\cdot{\bf H}^*\right]}.
\label{S73}
\end{equation}

In the last result one can observe that contributions involving the damping parameter $\gamma_0$, i.e., accounting for the material absorption effects, are $\pi/2$ out of phase with respect to that of free space. We can also see that this result reduces to the well known and widely studied expression for the optical chirality density in vacuum provided that $\gamma_0=0$, as it should be expected. Furthermore, at least in this simple example, it is interesting to note that the terms involving the current density arrange in such a way that only yield outcomes depending on $\text{Re}{\left[{\bf E}\cdot{\bf H}^*\right]}$. In this regard, it is straightforward to show that in general,
\begin{equation}
{\bf E}\cdot{\bf H}^*\equiv\frac{i}{\omega\mu_0\mu^*(\omega)}\left[{\bf E}\cdot\left(\nabla\times{\bf E}^*\right)\right]=\frac{i}{\omega\mu_0\mu^*(\omega)}\left[\text{Re}{\left[{\bf E}\cdot\left(\nabla\times{\bf E}^*\right)\right]}+\frac{1}{2}\nabla\cdot\left({\bf E}^*\times{\bf E}\right)\right].
\label{S74}
\end{equation}
Because we are assuming a nonmagnetic medium, the first term in the right-hand side of the latter equation always vanishes when taking the real part. On the other hand, the second term only contributes for nonlinear polarizations. Still, even for circularly polarized light, due to the spatial derivatives, such a contribution is in general zero, except in very special cases involving the spatial dependence of the complex field amplitude, e.g., when considering an oscillating electric dipole. 

It is worth remarking that the whole development for obtaining the correct expression of the optical chirality density in dispersive and lossy media relies on the underlying structure of the continuity equation [Eq. \eqref{S3}]. Indeed, according to this approach, we have specially focused on representing the right-hand side of Eq. \eqref{S29} in terms of total time derivatives. To look into this possibility we attempted to reproduce the treatment employed for the derivation of the energy density \cite{Loudon1970sm,Ruppin2002sm}. However, the occurrence of curl terms hinders this realization and special care must be taken with the residual terms so as to avoid meaningless results. Still, it is important to realize that, when combining the electric and magnetic contributions [Eqs. \eqref{S53} and \eqref{S56}, respectively] in free space, these terms completely cancel each other:
\begin{eqnarray}
\nonumber\frac{1}{2}\left[\boldsymbol{\mathcal{E}}\cdot\partial_t\left(\nabla\times\boldsymbol{\mathcal{D}}\right)+\boldsymbol{\mathcal{H}}\cdot\partial_t\left(\nabla\times\boldsymbol{\mathcal{B}}\right)\right]
\nonumber&=&\frac{1}{2}\left\{\varepsilon_0\partial_t\left[\boldsymbol{\mathcal{E}}\cdot\left(\nabla\times \boldsymbol{\mathcal{E}}\right)\right]+\varepsilon_0\mu_0\partial_t\boldsymbol{\mathcal{E}}\cdot\partial_t\boldsymbol{\mathcal{H}}+\mu_0\partial_t\left[\boldsymbol{\mathcal{H}}\cdot\left(\nabla\times \boldsymbol{\mathcal{H}}\right)\right]-\varepsilon_0\mu_0\partial_t\boldsymbol{\mathcal{H}}\cdot\partial_t\boldsymbol{\mathcal{E}}\right\}\\
&=&\frac{1}{2}\left\{\varepsilon_0\partial_t\left[\boldsymbol{\mathcal{E}}\cdot\left(\nabla\times \boldsymbol{\mathcal{E}}\right)\right]+\mu_0\partial_t\left[\boldsymbol{\mathcal{H}}\cdot\left(\nabla\times \boldsymbol{\mathcal{H}}\right)\right]\right\}=\partial_t\mathcal{C}_{\rm vacuum}.
\label{S75}
\end{eqnarray}
However, taking into account Eqs. \eqref{S59} and \eqref{S60}, the residual terms, $\partial_t\boldsymbol{\mathcal{E}}\cdot\left(\nabla\times\boldsymbol{\mathcal{E}}\right)$ and $\partial_t\boldsymbol{\mathcal{H}}\cdot\left(\nabla\times\boldsymbol{\mathcal{H}}\right)$, and the corresponding terms accounting for the light-matter interaction (given by the second and third summation in Eqs. \eqref{S55} and \eqref{S58}), must be generally interpreted as the loss rate of the field chirality density.
\subsection{B. Optical chirality density in dispersive and lossless media: Brillouin's approach}
We will now determine a closed expression for the optical chirality density by direct evaluation of the Fourier integrals. This procedure resembles that gives rise to the Brillouin formula for the energy density \cite{Landausm,Brillouinsm}, and then is constrained by the same prescriptions. For simplicity, hereinafter it will be assumed a linear, homogeneous and isotropic medium.

To start with, let us expand the left-hand side of Eq. \eqref{S53} by using the Fourier integrals as given in Eq. \eqref{S1}:
\begin{eqnarray}
\nonumber\frac{1}{2}\boldsymbol{\mathcal{E}}\cdot\partial_t\left(\nabla\times\boldsymbol{\mathcal{D}}\right)&=&\frac{1}{2}\int_{-\infty}^{+\infty}{{\bf E}(\omega')e^{-i\omega' t}d\omega'}\cdot \partial_t\left\{\nabla\times\left[\int_{-\infty}^{+\infty}{{\bf D}(\omega)e^{-i\omega t}d\omega}\right]\right\}\\
&=&\frac{-i}{2}\int_{-\infty}^{+\infty}{\int_{-\infty}^{+\infty}{\omega{\bf E}(\omega')\cdot\left[\nabla\times{\bf D}(\omega)\right] e^{-i\left(\omega'+\omega\right) t}d\omega'}d\omega}.
\label{S76}
\end{eqnarray}
According to Eq. \eqref{S59}, this is actually the time rate of change of optical chirality. The instantaneous distribution of the optical chirality density can then be obtained by integrating the latter expression over time:
\begin{eqnarray}
\nonumber\mathcal{C}^{\rm electric}(t')&=&\frac{-i}{2}\int_{-\infty}^{t'}{\int_{-\infty}^{+\infty}{\int_{-\infty}^{+\infty}{\omega{\bf E}(\omega')\cdot\left[\nabla\times{\bf D}(\omega)\right] e^{-i\left(\omega'+\omega\right) t}d\omega'}d\omega}dt}\\
&=&\frac{1}{2}\int_{-\infty}^{+\infty}{\int_{-\infty}^{+\infty}{\left[\frac{\omega}{\omega'+\omega}\right]{\bf E}(\omega')\cdot\left[\nabla\times{\bf D}(\omega)\right] e^{-i\left(\omega'+\omega\right) t'}d\omega'}d\omega},
\label{S77}
\end{eqnarray}
where it has been implicitly assumed fields tending sufficiently rapidly to zero as $t\to-\infty$. This is indeed a crucial assumption that necessarily restricts the applicability of this procedure to electromagnetic fields under the \textit{slowly varying amplitude approximation} \cite{Novotnysm}. In linear, homogeneous and isotropic media, $\nabla\times{\bf D}(\omega)=i\omega\varepsilon_0\mu_0\varepsilon(\omega)\mu(\omega){\bf H}(\omega)$, and then we have
\begin{equation}
\mathcal{C}^{\rm electric}(t)=\frac{i}{2c^2}\int_{-\infty}^{+\infty}{\int_{-\infty}^{+\infty}{\left[\frac{\omega^2\varepsilon(\omega)\mu(\omega)}{\omega'+\omega}\right]{\bf E}(\omega')\cdot{\bf H}(\omega) e^{-i\left(\omega'+\omega\right) t}d\omega'}d\omega}.
\label{S78}
\end{equation}
Likewise, performing the same calculation for the magnetic contribution it follows that
\begin{eqnarray}
\nonumber\mathcal{C}^{\rm magnetic}(t)&=&\frac{-i}{2c^2}\int_{-\infty}^{+\infty}{\int_{-\infty}^{+\infty}{\left[\frac{\omega^2\varepsilon(\omega)\mu(\omega)}{\omega'+\omega}\right]{\bf E}(\omega)\cdot{\bf H}(\omega')e^{-i\left(\omega'+\omega\right)t}d\omega'}d\omega}\\
&&+\frac{\mu_0}{2}\int_{-\infty}^{+\infty}{\int_{-\infty}^{+\infty}{\left[\frac{\omega\sigma(\omega)\mu(\omega)}{\omega'+\omega}\right]{\bf E}(\omega)\cdot{\bf H}(\omega')e^{-i\left(\omega'+\omega\right)t}d\omega'}d\omega}.
\label{S79}
\end{eqnarray}
Therefore, summing up the latter results we obtain that
\begin{equation}
\mathcal{C}^{\rm elec+magn}(t)=\frac{i}{4c^2}\int_{-\infty}^{+\infty}{\int_{-\infty}^{+\infty}{\left[\frac{\omega^2\varepsilon(\omega)\mu(\omega)\Xi(\omega',\omega)e^{-i\left(\omega'+\omega\right) t}+\omega'^2\varepsilon^*(\omega')\mu^*(\omega')\Xi^*(\omega',\omega)e^{i\left(\omega'+\omega\right) t}}{\omega'+\omega}\right] d\omega'}d\omega},
\label{S80}
\end{equation}
where $\Xi(\omega',\omega)={\bf E}(\omega')\cdot{\bf H}(\omega)-{\bf E}(\omega)\cdot{\bf H}(\omega')$. Note that in the above expression it has only been accounted the first term of the right-hand side of Eq. \eqref{S79}. By regarding monochromatic optical fields, it can be shown that
\begin{equation}
\Xi_{\rm av}(\omega',\omega)=\frac{i}{2}\text{Im}{\left[{\bf E}^*\cdot{\bf H}\right]}\left[\delta(\omega-\omega_0)\delta(\omega'+\omega_0)-\delta(\omega+\omega_0)\delta(\omega'-\omega_0)\right]=-\Xi^*_{\rm av}(\omega',\omega),
\label{S81}
\end{equation}
where there are only the terms contributing to the time average. Hence, substituting this into Eq. \eqref{S80} we find that the time-averaged optical chirality density is given by
\begin{equation}
\braket{\mathcal{C}^{\rm electric+magnetic}}_{\rm T}=\frac{1}{4c^2}\text{Im}{\left[{\bf E}\cdot{\bf H}^*\right]}\text{Re}{\left[\frac{\omega_0'^2\varepsilon(\omega_0')\mu(\omega_0')-\omega_0^2\varepsilon(\omega_0)\mu(\omega_0)}{\omega_0'-\omega_0}\right]}.
\label{S82}
\end{equation}
The latter result has a singular behavior when $\omega_0'=\omega_0$. To overcome this issue we should take the limit $\omega_0'\to\omega_0$, thus transforming the above expression into a derivative with respect to $\omega$:
\begin{equation}
\braket{\mathcal{C}^{\rm electric+magnetic}}_{\rm T}=\frac{1}{4c^2}\text{Re}{\left[\frac{d\left[\omega^2\varepsilon(\omega)\mu(\omega)\right]}{d\omega}\right]}\text{Im}{\left[{\bf E}\cdot{\bf H}^*\right]}=\frac{\omega}{2c^2}\text{Re}{\left[n(\omega)\tilde{n}(\omega)\right]}\text{Im}{\left[{\bf E}\cdot{\bf H}^*\right]},
\label{S83}
\end{equation}
where $n(\omega)=\sqrt{\varepsilon\mu}$ is the phase refractive index and $\tilde{n}(\omega)\equiv n+\omega[\partial n/\partial\omega]$ is the dispersion-modified group refractive index \cite{Bliokh2017sm}. A more elegant form for expressing the above result may be made regarding the phase and group velocities, $v_p(\omega)\equiv c/n$ and $v_g(\omega)\equiv c/\tilde{n}$, respectively:
\begin{equation}
\braket{\mathcal{C}^{\rm electric+magnetic}}_{\rm T}=\text{Re}{\left[\frac{\omega}{2v_p(\omega)v_g(\omega)}\right]}\text{Im}{\left[{\bf E}\cdot{\bf H}^*\right]},
\label{S84}
\end{equation}
thus showing that the stored optical chirality increases as the group and/or phase velocities tend to get smaller \cite{Boyd2009sm}.

In order to compare with the results derived following the Loundon's approach [Eq. \eqref{S64}], we should also consider the second integral of Eq. \eqref{S79}. This is carried out in the same way as the former calculations, and leads to
\begin{equation}
\braket{\mathcal{C}^{\rm current}}_{\rm T}=\frac{1}{8c^2}\text{Im}{\left[\frac{d\left[\omega^2\left(\varepsilon(\omega)-1\right)\mu(\omega)\right]}{d\omega}{\bf E}\cdot{\bf H}^*\right]}.
\label{S85}
\end{equation}
\subsection{C. Loundon's approach vs Brillouin approach: some additional remarks}
To complete the analysis, we now show that there exists certain condition under which the expression for the optical chirality density associated to the fields, exactly coincide for both the above approaches. Indeed, it is easy to see that the optical chirality density given in Eq. \eqref{S63} is equal to that expressed in Eq. \eqref{S83} (or Eq. \eqref{S84}) when
\begin{equation}
2\text{Re}{\left[\varepsilon(\omega)\mu(\omega)+\frac{\omega\mu(\omega)}{2}\frac{\partial \varepsilon(\omega)}{\partial\omega}+\frac{\omega\varepsilon(\omega)}{2}\frac{\partial \mu(\omega)}{\partial \omega}\right]}=\varepsilon_{\rm eff}\mu^*(\omega)+\varepsilon(\omega)\mu_{\rm eff}.
\label{S86}
\end{equation}
From the latter equation it is straightforward to observe that the imaginary part of the right-hand side must be zero. This translates into the following condition:
\begin{equation}
\chi''_n\left(1+\xi_n'\right)-\xi_n''\left(1+\chi_n'\right)+2\omega\chi_n''\xi_n''\left(\frac{1}{\tilde{\gamma}_n}-\frac{1}{\gamma_n}\right)=0,
\label{S87}
\end{equation}
which is fulfilled when $\gamma_n=\tilde{\gamma}_n=0$ for all $n$. At the same time, we can verify that this solution is indeed the correct one, just by substituting it into the real part of Eq. \eqref{S86}.
\begin{figure*}[t!]
	\includegraphics[width=0.75\linewidth]{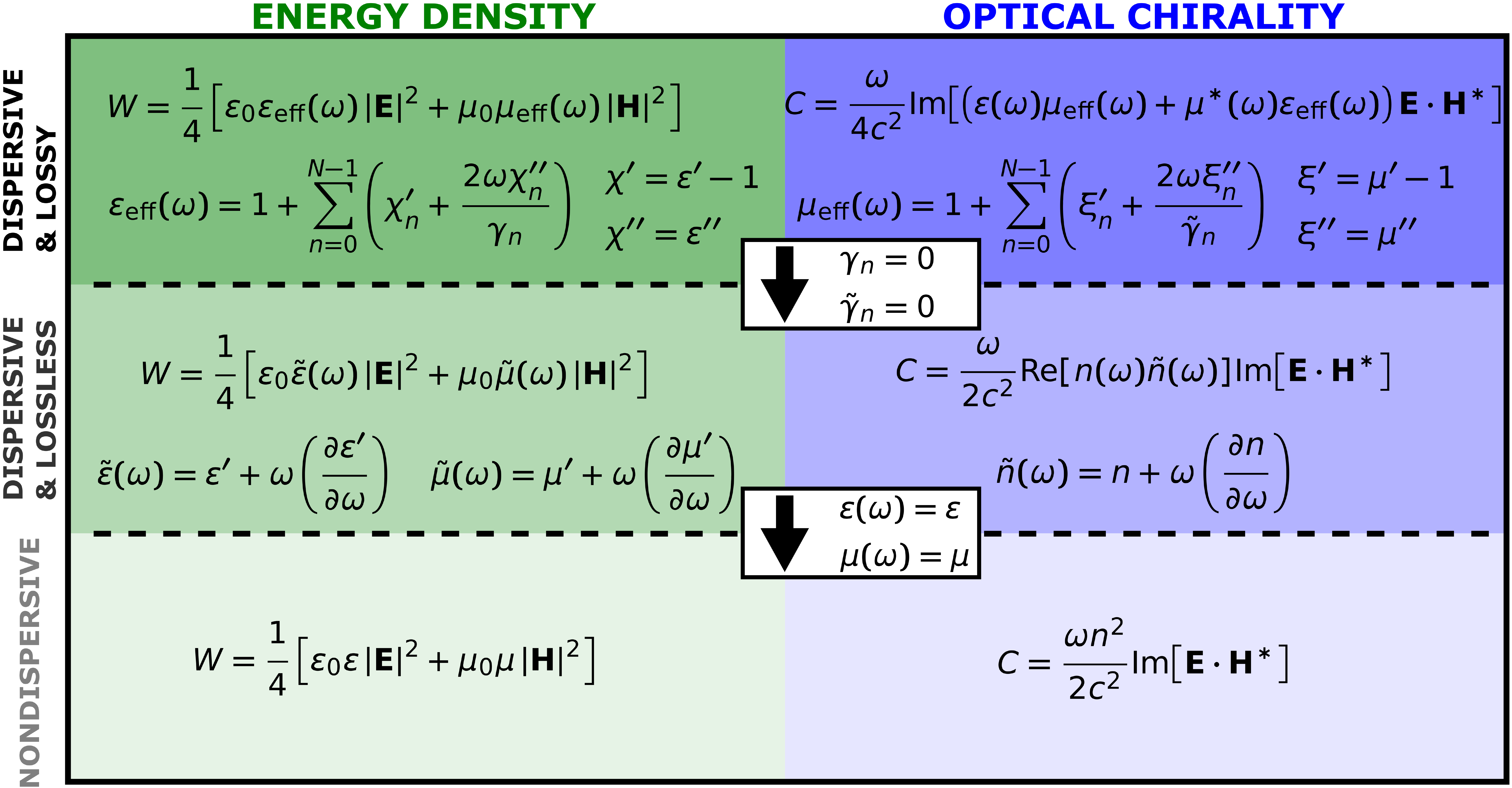}
	\caption{Schematic representation summarizing the main results showed in this supplementary material. For comparison, we also draw a parallel between the electromagnetic energy and the optical chirality densities derived by using different approaches.}
	\label{fig1sm}
\end{figure*}

Finally, we would like to highlight once again that the key point throughout this derivation relies on the underlying mathematical structure of the conservation laws [Eq. \eqref{S3}]. The classical approaches provided by Brillouin and Landau enable us to obtain a closed expression, but it is only valid under the slowly varying amplitude approximation, i.e., in a relatively narrow frequency range where the effects of material absorption are negligible. Therefore, Eqs. \eqref{S26} and \eqref{S83} are suitable for describing energy and chirality solely in lossless dispersive media. In order to further include the dissipation, we have to consider explicitly the material features as well. This is carried out by means of the corresponding dynamic equations characterizing the polarization and the magnetization fields of the medium, thereby introducing additional terms that must be explicitly included in the time derivative of Eq. \eqref{S3}. By doing so, we are able to find a more general definition for the energy and the optical chirality densities, thus capturing analytically the dispersion as well as the absorption processes. These results, given in Eqs. \eqref{S21}, \eqref{S22} and \eqref{S63}, are schematically summarized in Fig. \ref{fig1sm}. Importantly, departing from the top of the scheme, i.e., from the most general expressions, it is possible to arrive at the standard definitions, displayed in the intermediate and the lower panels of the scheme, by means of the corresponding approximations, thus progressively particularizing our general result. Hence, we can conclude that our approach leads to a consistent and meaningful definition for describing the optical chirality in dispersive and lossy media.
\subsection{D. Loss rate of the optical chirality in dispersive and lossy media}
For the sake of completeness, we will show the expressions for describing the loss (or gain) rate of the field chirality density in dispersive media. The source-like contribution associated with the energy conservation, $\boldsymbol{\mathcal{J}}\cdot\boldsymbol{\mathcal{E}}$, can be directly related to the power lost (or the work exerted) by the electromagnetic fields to the sources. However, the physical significance of the source-like terms associated with the optical chirality, $\boldsymbol{\mathcal{E}}\cdot\left(\nabla\times \boldsymbol{\mathcal{J}}\right)/2$, is not so obvious. This limits to some extent the concept of source (or sink) of optical chirality, and thus, hinders a proper interpretation of the loss rate associated with its conservation law \cite{Nienhuis2016sm,Corbaton2017sm}. Despite that, as pointed out above, these contributions are important because would enable us to get deeper insights into fundamental aspects regarding the interaction between chiral light with matter \cite{Schaferlingsm}. 

Leaving aside the difficulty surrounding the physical interpretation, from a mathematical point of view, the loss rate of the optical chirality can be derived taking into account the structure of the continuity equation [Eq. \eqref{S3}]. Then, according to the complete expression of the conservation law for the optical chirality given in Eq. \eqref{S65}, we will focus on the source-like terms $\mathcal{L}_{\rm chirality}^{\rm electric}$, $\mathcal{L}_{\rm chirality}^{\rm magnetic}$, and $\mathcal{L}_{\rm chirality}^{\rm current}$. For simplicity, we will consider a nonmagnetic medium, so that $\mathcal{L}_{\rm chirality}^{\rm magnetic}=0$. The electric contribution, $\mathcal{L}_{\rm chirality}^{\rm electric}$, is given by the second term of the right-hand side of Eq. \eqref{S54} together with the second and the third summation in Eq. \eqref{S55}. Once again, in order to compare these results with those obtained via the Fourier transform we have to calculate the time average by considering time harmonic fields. In this manner it can be demonstrated that the residual term of Eq. \eqref{S54} leads to
\begin{equation}
-\frac{\varepsilon_0}{2}\braket{\partial_t\boldsymbol{\mathcal{E}}\cdot\left(\nabla\times \boldsymbol{\mathcal{E}}\right)}_{\rm T}=\frac{\omega^2}{4c^2}\text{Re}{\left[\mu^*{\bf E}\cdot{\bf H}^*\right]}.
\label{S88}
\end{equation}
By summing this latter result and the time-average of the second and third summation in Eq. \eqref{S55} we obtain that
\begin{equation}
\braket{\mathcal{L}_{\rm chirality}^{\rm electric}}_{\rm T}=\frac{\omega^2}{4c^2}\left\{\text{Re}{\left[\varepsilon(\omega)\right]}\text{Re}{\left[\mu^*{\bf E}\cdot{\bf H}^*\right]}+\text{Im}{\left[\varepsilon(\omega)\right]}\text{Im}{\left[\mu^*{\bf E}\cdot{\bf H}^*\right]}\right\}.
\label{S89}
\end{equation}
On the other hand, assuming a linear, homogeneous and isotropic medium the current-like term is
\begin{equation}
\braket{\mathcal{L}_{\rm chirality}^{\rm current}}_{\rm T}=\frac{1}{2}\braket{\boldsymbol{\mathcal{E}}\cdot\left(\nabla\times \boldsymbol{\mathcal{J}}\right)}_{\rm T}=\frac{\omega^2}{4c^2}\left\{\text{Re}{\left[(\varepsilon-1)\right]}\text{Re}{\left[\mu^*{\bf E}\cdot{\bf H}^*\right]}+\text{Im}{\left[\varepsilon\right]}\text{Im}{\left[\mu^*{\bf E}\cdot{\bf H}^*\right]}\right\}.
\label{S90}
\end{equation}
Hence, the loss rate of the optical chirality in a nonmagnetic, dispersive and lossy medium is given by
\begin{equation}
\braket{\mathcal{L}_{\rm chirality}}_{\rm T}=\braket{\mathcal{L}_{\rm chirality}^{\rm electric}}_{\rm T}+\braket{\mathcal{L}_{\rm chirality}^{\rm current}}_{\rm T}=\frac{\omega^2}{2c^2}\left\{\text{Re}{\left[\varepsilon-\frac{1}{2}\right]}\text{Re}{\left[{\bf E}\cdot{\bf H}^*\right]}+\text{Im}{\left[\varepsilon\right]}\text{Im}{\left[{\bf E}\cdot{\bf H}^*\right]}\right\}.
\label{S91}
\end{equation}
\subsection{E. Loss rate of the optical chirality in dispersive and lossless media}
Since we are assuming a lossless medium, the only term accounting for chirality dissipation is the current-like term $\boldsymbol{\mathcal{E}}\cdot\left(\nabla\times \boldsymbol{\mathcal{J}}\right)/2$. Taking the Fourier transform we arrive at 
\begin{eqnarray}
\nonumber\frac{1}{2}\boldsymbol{\mathcal{E}}\cdot\left(\nabla\times\boldsymbol{\mathcal{J}}\right)&=&\frac{1}{2}\int_{-\infty}^{+\infty}{{\bf E}(\omega')e^{-i\omega't}d\omega'}\cdot\left(\nabla\times\left[\int_{-\infty}^{+\infty}{{\bf J}(\omega)e^{-i\omega t}d\omega}\right]\right)\\
\nonumber&=&\frac{1}{2}\int_{-\infty}^{+\infty}{\int_{-\infty}^{+\infty}{\sigma(\omega){\bf E}(\omega')\cdot \left[\nabla\times{\bf E}(\omega)\right]e^{-i\left(\omega'+\omega\right)t}d\omega'}d\omega}\\
&=&\frac{i\mu_0}{2}\int_{-\infty}^{+\infty}{\int_{-\infty}^{+\infty}{\omega\mu(\omega)\sigma(\omega){\bf E}(\omega')\cdot  {\bf H} (\omega)e^{-i\left(\omega'+\omega\right)t}d\omega'}d\omega}.
\label{S92}
\end{eqnarray}
By considering monochromatic optical fields, and integrating over frequencies we finally obtain
\begin{equation}
\frac{1}{2}\braket{\boldsymbol{\mathcal{E}}\cdot\left(\nabla\times\boldsymbol{\mathcal{J}}\right)}_{\rm T}=\frac{\omega^2}{4c^2}\left\{\text{Re}{\left[(\varepsilon-1)\right]}\text{Re}{\left[\mu^*{\bf E}\cdot{\bf H}^*\right]}+\text{Im}{\left[\varepsilon\right]}\text{Im}{\left[\mu^*{\bf E}\cdot{\bf H}^*\right]}\right\}.
\label{S93}
\end{equation}
As we can see, this result exactly coincides with that given in Eq. \eqref{S90}. However, in lossy media there are additional contributions that leads to the complete expression [Eq. \eqref{S91}] accounting for the sources or sinks of optical chirality.

\begin{thebibliography}{90}
\bibitem{footnote1sm} Notice that, strictly speaking, vacuum is the only nondispersive medium \cite{Jacksonsm,Novotnysm}. Still, dielectrics are often regarded in a simpler fashion as transparent (or nonabsorbing) and nondispersive material systems, or even as lossless dispersive media over some frequency bands for which absorption losses are negligibly small \cite{Landausm,Brillouinsm}.
\end{thebibliography}
\end{document}